\documentclass[twocolumn,aps,repreprint]{revtex4-1} 
\usepackage{amsmath,amssymb,graphicx,times,color,upgreek}
\usepackage[colorlinks, linkcolor=blue, citecolor=blue, urlcolor=blue, breaklinks=true]{hyperref}
\usepackage{amsmath}

\begin{document}

\title{Electro-optic entanglement source \\for microwave to telecom quantum state transfer}
\author{Alfredo Rueda}
\email{aruedasa@ist.ac.at}
\author{William Hease}
\author{Shabir Barzanjeh}
\author{Johannes M. Fink}
\email{jfink@ist.ac.at}

\affiliation{Institute of Science and Technology Austria, am Campus 1, 3400 Klosterneuburg, Austria}%
\date{\today}
\begin{abstract}

We propose an efficient microwave-photonic modulator as a resource for stationary entangled microwave-optical fields and develop the theory for deterministic entanglement generation and quantum state transfer in multi-resonant electro-optic systems. 
The device is based on a single crystal whispering gallery mode resonator integrated into a 3D microwave cavity. 
The specific design relies on a new combination of thin-film technology and conventional machining that is optimized for the lowest dissipation rates in the microwave, optical and mechanical domains.
We extract important device properties from finite element 
simulations 
and predict continuous variable entanglement generation rates on the order of a Mebit/s for optical pump powers of only a few tens of microwatt. We compare the quantum state transfer fidelities of coherent, squeezed and non-Gaussian cat-states for both teleportation and direct conversion protocols under realistic conditions. Combining the unique capabilities of circuit quantum electrodynamics with the resilience of fiber optic communication could facilitate long distance solid-state qubit networks, new methods for quantum signal synthesis, quantum key distribution, and quantum enhanced detection, as well as more power-efficient classical sensing and modulation.\end{abstract}
               
\maketitle

\section{Introduction}
The development of superconducting quantum processors has seen remarkable progress in the last decade \cite{schoelkopf_wiring_2008,Wendin_2017}, but long distance connectivity remains an unsolved problem. Coherent interconnects between superconducting qubits are currently restricted to an ultra-cold environment, which offers sufficient protection from thermal noise \cite{Kupiers2018,Chou2018}. A hybrid quantum network that combines the advanced control capabilities and the high speed offered by superconducting quantum circuits, with the robustness, range \cite{Liao2018} and versatility \cite{Maring17} of more established quantum telecommunication systems appears as the natural solution. 
Entanglement between optical and microwave photons is the key ingredient for distributed quantum computing with such a hybrid quantum network and would pave the way to integrate advanced microwave quantum state synthesis capabilities \cite{Hofheinz2009, Eichler2011a, Vlastakis2013} with existing optical quantum information protocols~\cite{Braunsteinvanloock, weedbrook2012} such as quantum state teleportation \cite{Furusawa706,Lee330} and secure remote quantum state preparation~\cite{Laurat03,pogorzalek2019}. 

Electro-optomechanical systems stand out as the most successful platforms to connect optical and microwave fields near losslessly and with minimal added noise  \cite{Andrews2014,higgin}. 
Very recently it has been shown that mechanical oscillators can also be used to deterministically generate entangled electromagnetic fields
\cite{Barzanjeh2019}. 
Mechanical generation of microwave-optical entanglement has been proposed \cite{Genes2008b, Stannigel2010, Barzanjeh2011a, shab1, Wang2013, Tian2013, Zhong2019} but an experimental realization remains challenging. Low frequency mechanical transducers typically suffer from added noise and low bandwidth, while high frequency piezoelectric devices 
require sophisticated wave matching and new materials, which so far results in low total interaction efficiencies \cite{Bochmann2013, Forsch2018, Shao2019}, comparable to magnon-based interfaces \cite{hisatomi}.

Cavity electrooptic (EO) modulators are another proposed candidate \cite{Matsko2007, tsang_cavity_2010, tsang_cavity_2011, Javerzac-Galy2016, Soltani2017} to coherently convert photons, or to effectively generate entanglement between microwave and optical fields, employing the Pockels effect and without the need for an intermediary oscillator. Here, a material with a large and broadband nonlinear polarizability $\chi^{(2)}$ is shared between an optical resonator and the capacitor of a microwave cavity \cite{cohen_microphotonic_2001-1, ilchenko_whispering-gallery-mode_2003, strekalov_efficient_2009, strekalov_microwave_2009,botello_sensitivity_2018}, a platform that has recently been used for efficient photon conversion with bulk \cite{Rueda:16} and thin film crystals \cite{FanZouChengEtAl2018}. 

In this paper, we propose a multi-resonant whispering gallery mode (WGM) cavity electro-optic modulator who's free spectral range matches the microwave resonance frequency. It is optimized for optimal performance at ultra-low temperatures, in particular with respect to unwanted optical heating and thermal occupation of the microwave mode. We minimize the necessary optical pump power by maximizing the optical quality factor using a millimeter sized and mechanically polished bulk single crystal disk resonator \cite{reviewstrekalov}. 
Compared to nano- and micron-scale modulators its large size and surface area should facilitate a more efficient coupling to the cold bath and its large heat capacity is expected to result in slow heating rates in pulsed operation schemes. Compared to previous work \cite{Rueda:16} the disk is clamped in the center to avoid disk damage, air gaps and to minimize potential piezoelectric clamping losses. Importantly, finite-element modeling shows that a sufficient mode overlap and bandwidth at moderate pump powers can still be achieved using a combination of lithographically defined thin-film superconducting electrodes together with carefully shaped WGM disc cross-sections.

In the main part of the paper we develop the theory to analytically predict the entanglement properties under realistic conditions such finite temperature and asymmetric waveguide couplings. We show that it is feasible to deterministically generate MHz bandwidth continuous variable (CV) entanglement between the outputs of a pumped optical and a cold microwave resonator via spontaneous parametric downconversion (SPDC). We also present its performance for direct conversion-based and teleportation-based communication, as quantified by the quantum state transfer fidelities for a set of typical quantum states. Our results indicate that the proposed entangler could serve as a repeater node to enable long distance hybrid quantum networks \cite{Muralidharan2016}. The developed theory results are applicable to any triply-resonant electro-optic transducer implementation.

\section{SYSTEM}
\subsection{Hamiltonian of the system} 
\begin{figure}[t!]
	\centering
		\includegraphics[width=0.8\columnwidth]{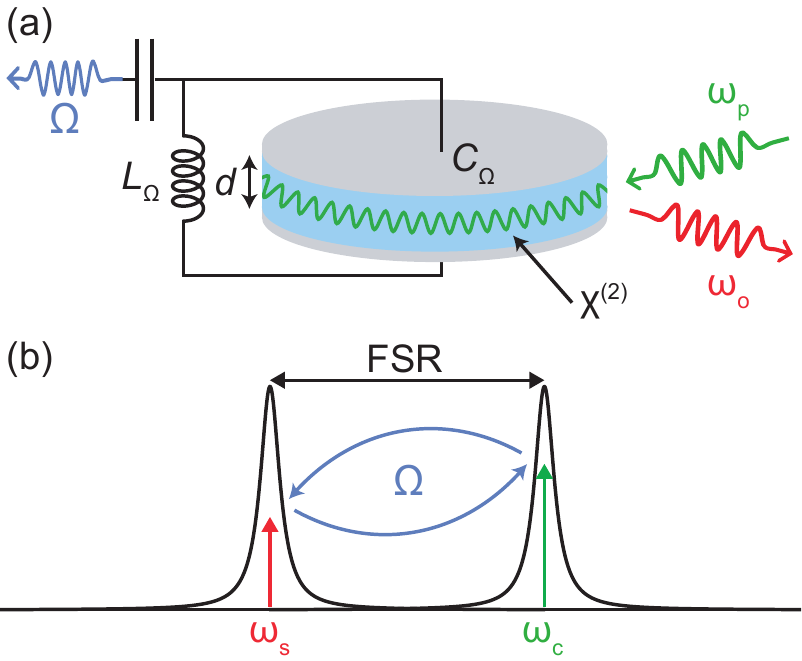}
			\caption{Schematic representation of the cavity electro-optic modulator. (a) An optical WGM resonator with $\chi^{(2)}$ nonlinearity is confined between two metallic electrodes forming the capacitance $C_\Omega$ of a LC microwave resonator with resonance frequency $\Omega=1/\sqrt{L_\Omega C_\Omega}$. An incident optical pump field at $\omega_p$ is down-converted to two outgoing correlated microwave-optical fields $\Omega$ and $\omega_o$. (b) The power spectral density of the optical resonator. The two shown modes of the WGM resonator correspond to the central and Stokes sideband modes with resonance frequencies $\omega_c$ and $\omega_{s}$. Efficient microwave-optical interaction requires matching $\Omega$ with the free spectral range (FSR) of the optical mode. Here, the optical resonator is coherently pumped at resonance frequency $\omega_p=\omega_c$ and the output of the optical resonator is measured at the Stokes-Sideband frequency $\omega_o=\omega_s$. }
	\label{simple}
\end{figure}

As shown schematically in Fig.~\ref{simple}, we consider a WGM cavity electro-optic modulator containing a $\chi^{(2)}$ nonlinear medium that generates a nonlinear interaction between a single microwave cavity mode with frequency $\Omega$ and two modes of the WGM optical resonator corresponding to the central and the Stokes sideband mode with resonance frequencies $\omega_c$ and $\omega_s$, respectively. Such a single sideband situation can be achieved by making use of mode couplings of different polarization that lead to an asymmetry of the free spectral range of the WGM \cite{Rueda:16}. The total Hamiltonian describing the system is $\hat H=\hat H_0+\hat H_{\mathrm{int}}$ in which the free energy Hamiltonian is \cite{ilchenko_whispering-gallery-mode_2003,Matsko2007,tsang_cavity_2010,tsang_cavity_2011,Rueda:16}
\begin{equation}
\hat{H}_0=\hbar\omega_c a_c^{\dagger}a_c+\hbar\omega_{s}a_s^{\dagger}a_s+\hbar\Omega a_\Omega^{\dagger}a_\Omega,
\end{equation}
and the interaction Hamiltonian is
\begin{equation}
\hat{H}_\text{int}= g (a_{\Omega}+a_{\Omega}^{\dagger})(a_{c}^{\dagger}+a_{s}^{\dagger})(a_{c}+a_{s}), \label{hamscat0}
\end{equation}
where $\hat a_c$, $\hat a_{s}$ are the annihilation operators of the central and Stokes sideband modes of the optical resonator, respectively, while $\hat a_\Omega$ is the annihilation operator of the microwave cavity and $g$ describes the coupling strength between the microwave and the two optical modes. Moving to the interaction picture with respect to $\hat H_0$ and setting $\Omega=\omega_c-\omega_s$, the system Hamiltonian reduces to
\begin{equation}
\hat{H}= g (\hat{a}_c^\dagger\hat{a}_s \hat{a}_{\Omega}+\hat{a}_{\Omega}^\dagger\hat{a}_s^\dagger \hat{a}_c). \label{hamscat}
\end{equation}
The second part of this Hamiltonian describes a three-wave mixing process during which an optical photon with frequency $\omega_s$ and a microwave photon with frequency $\Omega$ are generated by annihilating an optical photon with frequency $\omega_c$. 

The coupling strength $g$ is determined by the spatial mode overlap of the electric fields 
$E_k = \sqrt{\hbar \omega_k / (2 \epsilon_k V_k)} \psi_k(r,\theta,\phi)$ and the $\chi^{(2)}$ nonlinearity of the material \cite{ilchenko_whispering-gallery-mode_2003,Rueda:16}:
\begin{equation}
g=2\epsilon_0 \chi^{(2)}  \sqrt{\frac{\hbar \omega_s  \omega_c \Omega}{8\epsilon_s\epsilon_c\epsilon_{\Omega}V_cV_sV_{\Omega}}} \int dV \psi_s^* \psi_c\psi_{\Omega}.\label{eq:defgg}
\end{equation}
where $\epsilon_0$ is the vacuum permittivity, $\psi_k(r,\theta,\phi)$ the field distribution functions, $\epsilon_k$ and $V_k$ are the relative permittivity and mode volume corresponding to mode $k$ with $k = s, c, \Omega$, respectively. The field distributions can be written in terms of the cross section $\Psi_k(r,\theta)$ and azimuthal distribution $e^{-im_k\phi}$ as  $\psi_k(r,\theta,\phi)=\Psi_k(r,\theta)e^{-im_k\phi}$. The integral over the azimuthal variable $\phi$ is nonzero only if the relation $m_c=m_s+ m_{\Omega}$ is fulfilled. This condition, known as phase matching or angular momentum conservation, returns a real value of the coupling constant $g$ presented in Eq (\ref{eq:defgg}). 

We can linearize the Hamiltonian in Eq. (\ref{hamscat}) by limiting our analysis to the case where the center mode of the optical cavity is pumped resonantly by a strong coherent field at frequency $\omega_p=\omega_c$. In this condition the optical mode $\hat a_c$ can be treated as classical complex number $\alpha_p=\langle \hat a_c\rangle$ and the linearized Hamiltonian becomes
 \begin{equation}
\hat{H}=\hbar \alpha_p g (\hat{a}_o\hat{a}_{\Omega}+\hat{a}_{\Omega}^\dagger\hat{a}_{o}^\dagger). \label{hamsqu}
\end{equation}
Here for simplicity we renamed the optical mode $\hat a_s\rightarrow\hat a_o$. The above Hamiltonian describes a parametric down-conversion process that is responsible for entangling the microwave mode $\Omega$ with the optical mode $\omega_o$. In a lossless system, this interaction could lead to an exponential growth of the energy stored in both modes and consequently lead to photon amplification of each mode.

\subsection{Device implementation}
The proposed system is based on a 3D-microwave cavity enclosing a mm-sized LiNbO$_3$ WGM-resonator with major radius $R$ operating at millikelvin temperature. At optical wavelengths, these mechanically polished resonators offer material-limited internal quality factors $Q_{i,o}\gtrsim3.3\times10^8$ \cite{Leidinger:15} and strong lateral confinement, presented by the optical mode cross-section $\Psi_k(r,\theta)$, on the order of tens of $\mu m^2$  \cite{RuedaSanchez2018}. In the microwave regime LiNbO$_3$ exhibits an internal quality factor  $Q_{i,\Omega}\gtrsim10^4$ in the X-band at millikelvin temperatures \cite{goryachev_single-photon_2015} and a high electro-optic coefficient $r_{33}=31$ pm/V at 9 GHz \cite{Weis1985,wong2002properties}.

\begin{figure}[t]
	\centering
		\includegraphics[width=0.50\textwidth]{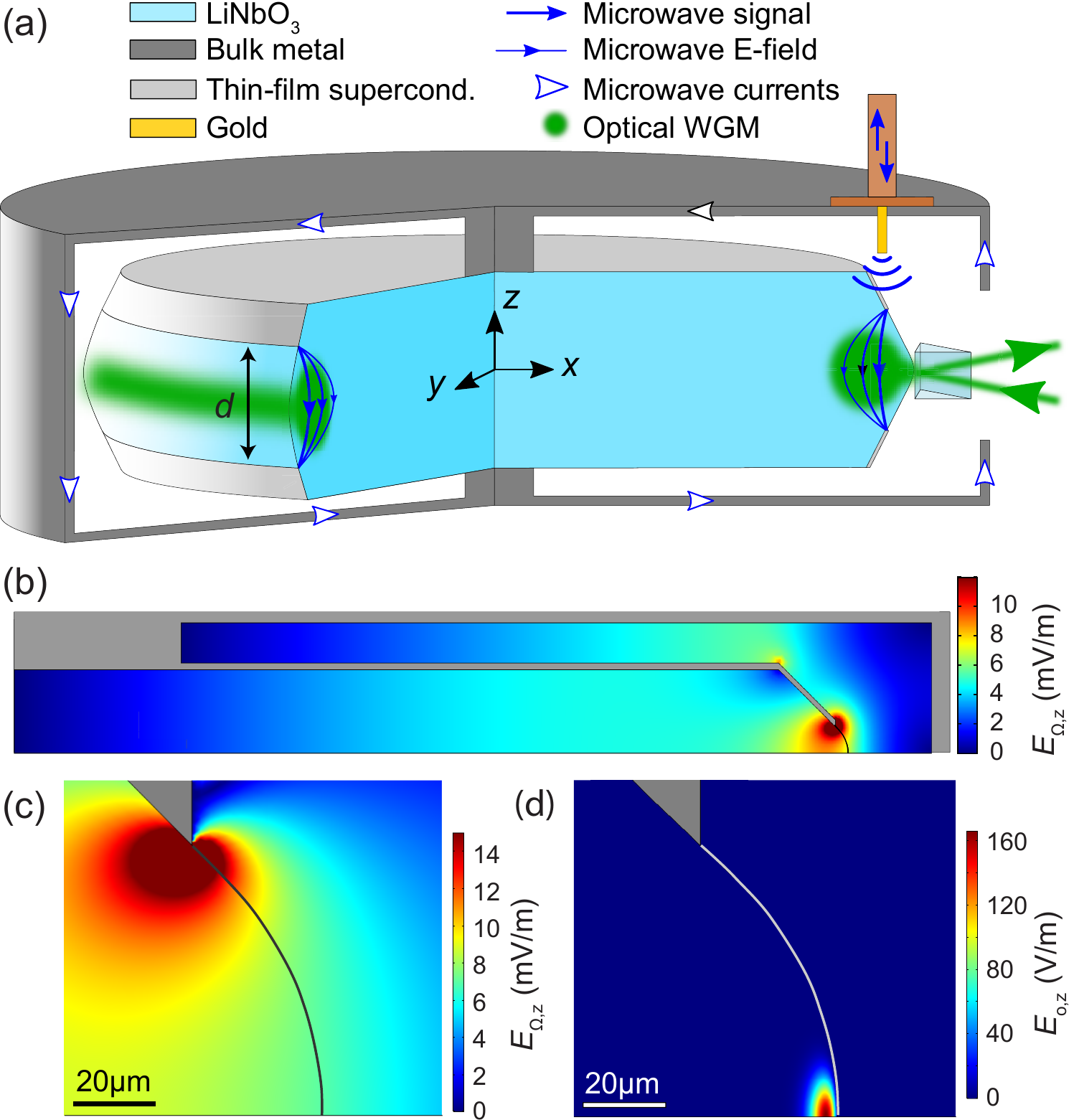}
			\caption{Device implementation of the proposed cavity electro-optic modulator.  (a) A monolithic LiNbO$_3$ optical resonator is incorporated inside a metal microwave cavity. The optical resonator is coated with a thin film superconductor that defines the capacitor gap $d$ and confines the microwave electric field at the resonator's rim. (b) To scale: microwave single photon electric field distribution $E_{\Omega,z}$ along the $z$-axis. Only a quarter of the resonator is shown for symmetry reasons. (c) Enlarged view of the electric field distribution  $E_{\Omega,z}$ at the resonator's rim. (d) Enlarged view of the electric field distribution of the optical mode  $E_{o,z}$ along the $z$ direction.}
	\label{setupsim}
\end{figure}

The large wavelength $\lambda_\Omega\gtrsim R$ of the microwave field causes considerable reduction of the spatial optical-microwave mode overlap, leading to a small microwave-optical mode coupling $g$. The proposed system tackles this problem by coupling the optical resonator to a metallic cavity. This hybrid device involves a monolithic LiNbO$_3$ resonator clamped at the center of a microwave cavity by two thin rods machined for example from aluminum or copper as depicted in Fig.~\ref{setupsim}(a). The LiNbO$_3$ resonator is coated with a thin film of superconductor such as Al or NbTiN 
forming the upper and lower electrodes of a capacitor for the microwave cavity. 
The thin film electrodes can be realized by evaporating metal on the full resonator's surface followed by optical lithography on the resonator's rim. The photoresist is developed and the unprotected thin metal band is etched. 
An interesting feature of this resonator fabrication process is the possibility to vary the gap size $d$ between the upper and lower electrodes independent of the disk thickness. Gaps from $1$ mm down to 10  $\mu$m are feasible by adjusting the focus of the lithography laser. This results in a strong confinement of the microwave electric field at the resonator's wedge-shaped rim, enhancing the mode overlap between the optical and microwave mode as shown in Fig.~\ref{setupsim}(b)-(d), and increasing the coupling constant $g$ (see Eq.~(\ref{eq:defgg})). In addition, the enclosing cavity offers a degree of freedom to control the microwave mode's spatial distribution $\psi_\Omega(\vec{r})$, the microwave resonance frequency $\Omega$ and its coupling to a microwave coaxial waveguide $\kappa_{e,\Omega}$.

To achieve optical-microwave mode interaction the energy and azimuthal momentum conservations must be fulfilled. For this system, we use and isolate two neighboring optical modes with angular number $m_s=m$ and $m_c=m+1$, spectrally separated by a free spectral range (FSR) as experimentally shown in \cite{Rueda:16}. 
The energy conservation is fulfilled by matching the microwave mode frequency $\Omega$ to the optical FSR.  Additionally, the microwave mode field distribution must have one oscillation around the resonator's rim ($m=1$) to fulfill the angular momentum conservation.  We assume the center frequency of the WGM resonator with mode number $m_c=m+1$ is coherently pumped via the evanescant coupling via a dielectric prism which also serves as the out-coupling port for the created Stokes-sideband with mode number $m_s=m$. On the microwave side, a pin coupler can be used to couple the microwave photons into a coaxial waveguide as depicted in Fig.~\ref{setupsim}(a). Here we work with a WGM resonator with an optical FSR of $9$~GHz, corresponding to the typical frequency range of superconducting qubits and read-out resonators.

\subsection{Numerical analysis of the system}
Figure \ref{setupsim} (c) and (d) show the numerical simulation of the electric field distribution of the microwave and optical modes, respectively. The microwave electric field is constant in the region enclosing the optical field. We simulate a $z$-cut LiNbO$_3$ WGM-resonator with the major radius $R=2.5$ mm, hight $H=0.5$ mm, and side curvature $R_c=0.1$ mm, enclosed by a cylindrical microwave cavity with diameter 5.5 mm and 1.3 mm hight. The optical WGM cross-section (FWHM) is analytically calculated to be $7.6\times17.8$ $\mu m^2$  \cite{RuedaSanchez2018}. For the electric fields along the $\hat{z}$ direction, the integral term in Eq.~(\ref{eq:defgg}) results in \cite{Rueda:16}:

\begin{equation}
g=\frac{1}{4\sqrt{2}}\cdot n_e^2 \cdot \omega_p \cdot r_{33} \cdot E_{\Omega,z}(\vec{r}_o), \label{qwithtroot}
\end{equation}

where $n_e$ is the extraordinary optical refractive index of LiNbO$_3$ and  $E_{\Omega,z}(r_o)$ is the $z$-component of the single photon microwave electric field at the position $\vec{r}_o$ of the optical mode. The $1/\sqrt{2}$ correction term is due to the nature of the microwave stationary wave, which can be seen as two contra-propagating waves, one of which only propagates opposite to the optical mode and therefore does not interact with it.

\begin{figure*}[t]
	\centering
		\includegraphics[width=0.85\textwidth]{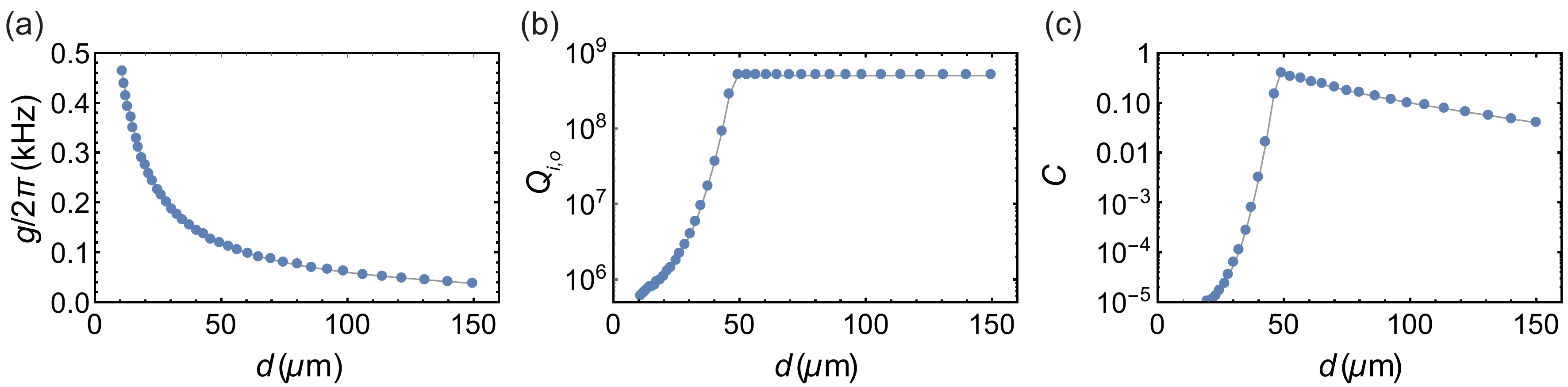}
	\caption{Simulated device parameters as a function of the gap size $d$ between the two thin-film metal electrodes.  (a) The microwave-optical coupling rate $g$, (b) the intrinsic optical quality factor $Q_{i,o}$,  and (c) the multi-photon cooperativity $C$ for $P_p=10$ $\mu$W as function of the electrodes gap size $d$ at 10 mK. }
	\label{EG}
\end{figure*}

Figure ~\ref{EG}(a) shows the simulated microwave-optical coupling rate $g$ as a function of the electrodes gap size $d$. From a parametric fit to the simulated values, we find the dependency of coupling rate $g$ to the gap size $d$ scales with $g\sim d^{-0.8}$. Figure~\ref{EG}(b) shows the internal quality factor of the optical resonator $Q_{i,o}$ versus the gap size $d$. By decreasing the gap size, $Q_{i,o}$ decreases exponentially because the optical mode has a Gaussian envelope $\Psi_k\sim\exp(-0.5\cdot z^2/\sigma_z^2)$ with $\sigma_z=7.6\, \mu$m along the $\hat{z}$-axis. For this simulation we consider aluminum electrodes, which exhibit a large imaginary index at optical frequencies \cite{Aluimag,Rakic:95} and therefore impose an optical loss for small electrode distances. Additionally, optical photons can break the cooper-pairs in the superconducting electrodes, degrading the quality factor of the microwave cavity. Therefore, it is desirable to reduce the spatial overlap between the optical mode and the superconducting electrodes and reduce the optical surface scattering. For moderately sized gaps the optical quality factor reaches the limit of $Q_{i,o}\sim5\times10^8$, which is backed by our experimental results at room temperature without the metal electrodes and expected to be the material-limited absorption of LiNbO$_3$.

We have also carried out characterization measurements of the microwave properties of the proposed system shown in Fig.~\ref{setupsim} using an aluminum cavity and thin-film aluminum metallization which yielded values $Q_{i,\Omega}\sim 3\times10^3$ for a clamping rod diameter of 0.5 mm. This value, which we use for our further modeling, is at least a factor 4 below the reported material limit of LiNbO$_3$ and we attribute this to other loss sources such as frequency dependent defect states \cite{goryachev_single-photon_2015}, piezoelectric mechanical \cite{Nguyen2013}, cavity seam \cite{Brecht16} and surface losses \cite{werner11} .

The multi-photon cooperativity 
$C=\frac{4 n_p g^2}{\kappa_o \kappa_\Omega}$ 
is the figure of merit in electro-optic systems. Here 
$n_p=|\alpha_p|^2=\frac{4\eta_o}{\kappa_o} \frac{P_p}{\hbar \omega_o}$ 
is the intra-cavity photon number due to the resonant optical pump power $P_p$, 
$\kappa_{\Omega(o)}=\kappa_{e,\Omega(o)}+\kappa_{i,\Omega(o)}$ is total loss of the microwave (optical) cavity while 
$\kappa_{e,\Omega(o)}$ and $\kappa_{i,\Omega(o)}$ are the extrinsic and the intrinsic damping rates of the microwave (optical) cavity, respectively. 
Here we defined the normalized cavity to waveguide coupling strength as 
$\eta_{\Omega(o)}=\kappa_{e,\Omega(o)}/\kappa_{\Omega(o)}$ of the microwave (optical) resonator. Under the critically coupled condition $\eta_\Omega=\eta_o=1/2$ the cooperativity is maximized for a given pump power
\begin{eqnarray} \label{coopera}
C = \frac{P_p g^2  Q_{i,o}^2  Q_{i,\Omega}}{\hbar\omega_p^3\Omega},
\end{eqnarray}
with the intrinsic quality factor $Q_{i,\Omega(o)}=\Omega(\omega_o)/\kappa_{i,\Omega(\omega_o)}$ of the microwave (optical) mode.
In Fig.~\ref{EG}(c) we plot the microwave-optical cooperativity $C$ as a function of the electrode gap size $d$ and for a fixed optical pump power of $10\,\mu$W. This plot shows that the cooperativity increases by decreasing the gap size and it reaches its maximum value at $d\sim50\,\mu$m where $Q_{i,o}$ starts to saturate due to material absorption. To reach strong multi-photon microwave-optical interaction requires a cooperativity close to 1, which can be achieved by increasing the optical pump power to $P_{p}=25.4\,\mu$W. 

It is important to note that this is lower than the cooling power of commercial cryostats at about 30 mK and that in practice only a small fraction of it would be dissipated into the cold stage of the dilution refrigerator, while the majority of the pump field is out-coupled together with the generated signal via an optical fiber e.g. by using a diamond prism with a basis angle of $63.5^\circ$. Nevertheless, in the following we will also consider the situation when the EO modulator is operated at the still plate at 800 mK and connected to a cold superconducting circuit at a few mK via a low-loss superconducting waveguide. The still stage of a modern dilution refrigerator offers cooling powers of at least 20 mW and the higher temperature offers higher thermal conductivities to connect the modulator more efficiently to the cold bath. Table~\ref{tab1} summarizes the full set of system parameters that will be used in the following unless otherwise stated. 

\begin{table}[h!]
\begin{ruledtabular}
\begin{tabular}{ccccccc}
$\Omega/2\pi\,$&$\omega_o/2\pi\,$&$g/2\pi\,$&$Q_{i,\Omega}$&$Q_{i,o}$&$\eta_\Omega$&$\eta_o$\\
\hline
 9 GHz & 193.5 THz& 119 Hz& $3\times10^3$ & $5\times10^8$ & 0.8 & 0.5
\end{tabular}
\end{ruledtabular}
\caption{Reference values for the proposed system based on simulation ($\Omega$, $\omega_o$ and $g$) and characterization measurements of the system ($Q_{i,\Omega}$ and $Q_{i,o}$). For generality we chose an asymmetric coupling situation $\kappa_\Omega > \kappa_o$ and $\eta_\Omega > \eta_o$. 
The necessary pump power to achieve $C=1$ in this asymmetrically and over-coupled configuration is $P_{p,C=1}=63.9\,\mu$W. }\label{tab1}
\end{table}

\section{System Dynamics}
In this section, we study the quantum dynamics of the proposed electro-optic modulator system presented in the previous section. We specifically focus on the conditions under which one can efficiently correlate and entangle optical and microwave fields using electro-optic interaction.  
The dynamics of the system can be fully described using the quantum Langevin treatment in which we add the damping and noise terms to the Heisenberg equations for the
system operators associated with Eq.(\ref{hamsqu}). The resulting quantum Langevin equations for the intra-cavity optical and microwave modes are
\begin{subequations}\label{threemodesDFG}
\begin{eqnarray}
\dot{\hat{a}}_{\Omega}&=&-iG \hat{a}^\dagger_{o}-\frac{\kappa_{\Omega}}{2} \hat{a}_{\Omega}+\sqrt{\kappa_{e,\Omega}}\hat{a}_{e,\Omega}+\sqrt{\kappa_{i,\Omega}}\hat{a}_{i,\Omega},\\
\dot{\hat{a}}_{o}&=&-iG \hat{a}^\dagger_{\Omega}-\frac{\kappa_{o}}{2} \hat{a}_{o}+\sqrt{\kappa_{e,o}}\hat{a}_{e,o}+\sqrt{\kappa_{i,o}}\hat{a}_{i,o},
\end{eqnarray}
\end{subequations}
where $G=\sqrt{n_p}ge^{i\phi_p} $ is the multi-photon interaction rate
and $\phi_p$ the phase of the pump. We also introduce the zero-mean microwave (optical) input noises given by $\hat{a}_{e,\Omega(o)}$ and $\hat{a}_{i,\Omega(o)}$, obeying the following correlation
functions
\begin{subequations}
\begin{align}
\langle \hat a_{k,\Omega(o)}
^{\dagger}(t) \hat a_{k,\Omega(o)}(t^{\prime})\rangle&=\bar n_{\Omega(o)}^k\delta(t-t^{\prime}),\\
\langle \hat a_{k,\Omega(o)}(t) \hat a_{k,\Omega(o)}
^{\dagger}(t^{\prime})\rangle & = (\bar n_{\Omega(o)}^k+1)\delta(t-t^{\prime}),
\end{align}
\end{subequations}
with $k = e,i$ where $ \bar n_{\Omega(o)}^e$ and $ \bar n_{\Omega(o)}^i$ are the equilibrium mean thermal
photon numbers of the microwave (optical) fields. 

\begin{figure*}
	\centering
		\includegraphics[width=0.75\textwidth]{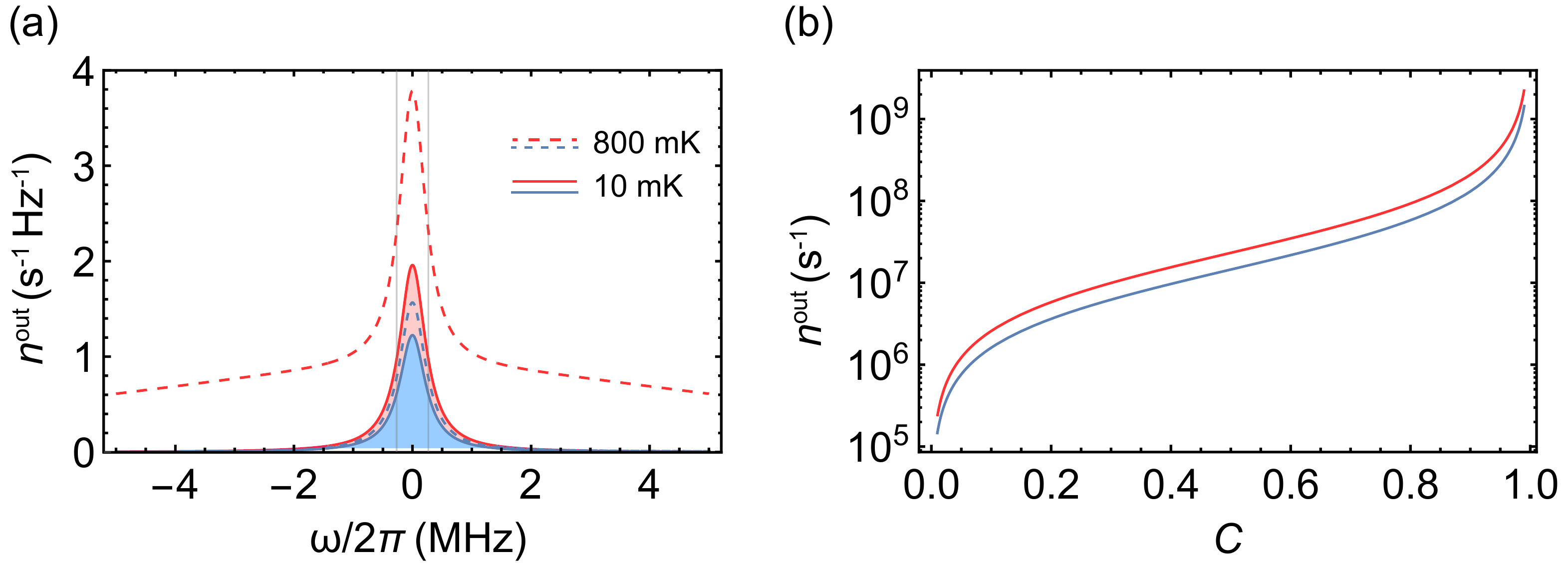}
\caption{Output photon numbers of the microwave and optical resonator.
(a) Output photon number spectral density at two bath temperatures $T_b=10$~mK (solid lines) and $T_b=800$~mK (dashed lines) of the microwave (blue) and optical resonator (red) for the values given in Table \ref{tab1} and $C=0.3$. 
(b) Total integrated output photon flux of the optical resonator (blue) and microwave cavity (red) with respect to the pump power dependent cooperativity $C$ at $T_b=10$~mK.}\label{fig:photonumber}
\end{figure*}

The equations (\ref{threemodesDFG}) describe the dynamics of the system and  reveal the origin of the optical-microwave intra-cavity correlation, which arises from the cross dependency of microwave operator $\hat a_\Omega$ on the optical mode operator $\hat a_o$, and vice versa. However, in this paper we are interested in generating nonclassical correlation and entanglement between itinerant electromagnetic modes, which can be calculated using the standard input-output theory \cite{Gardiner2004}. We first solve the Eqs. (\ref{threemodesDFG}) by moving to the Fourier domain to obtain the microwave and optical resonator variables. Then, substituting the solutions of Eqs. (\ref{threemodesDFG}) into the corresponding input-output formula for the cavities' variables, i.e., $\hat a_{\Omega(o)}^\text{out} = \sqrt{\kappa_{e,\Omega(o)}} \hat a_{\Omega(o)}  -\hat a_{e,\Omega(o)}$, we obtain
\begin{equation}
\hat{\text{S}}^\text{out}(\omega)=\textbf{D}(\omega)\cdot \hat{\text{S}}^\text{in}(\omega) \label{FDC}
\end{equation}
where $\hat{\text{S}}^\text{out}(\omega)=[a_o^\text{out}(\omega), a_\Omega^{\text{out}\dagger}(-\omega)]^\text{T}$ is the output field matrix. The transformation matrix $\textbf{D}(\omega)$ is given by \cite{tsang_cavity_2011}:
\begin{widetext}
\begin{footnotesize}
\begin{equation}
\textbf{D}(\omega) = M(\omega)^{-1}\begin{bmatrix}
(i\omega+\frac{\Delta\kappa_o}{2})(-i\omega+\frac{\kappa_\Omega}{2})+|G|^2 & \sqrt{\kappa_{e,o}\kappa_{i,o}}(-i\omega+\frac{\kappa_\Omega}{2})  &-iG\sqrt{\kappa_{e,o}\kappa_{e,\Omega}} & -iG\sqrt{\kappa_{e,o}\kappa_{i,\Omega}}  \\
iG^*\sqrt{\kappa_{e,\Omega}\kappa_{e,o}} & iG^*\sqrt{\kappa_{e,\Omega}\kappa_{i,o}} & (i\omega+\frac{\Delta\kappa_\Omega}{2})(-i\omega+\frac{\kappa_o}{2})+|G|^2 & \sqrt{\kappa_{e,\Omega}\kappa_{i,\Omega}}(-i\omega+\frac{\kappa_0}{2})  \label{matrix}
\end{bmatrix},
\end{equation}
\end{footnotesize}
\end{widetext}
with $M(\omega)=(-i\omega+\kappa_o/2)(-i\omega+\kappa_\Omega/2)-|G|^2$,  $\Delta\kappa_{\{o,\Omega\}}=\kappa_{e,\{o,\Omega\}}-\kappa_{i,\{o,\Omega\}}$ and $\hat{\text{S}}^\text{in}(\omega)$ is the input noise matrix $[ \hat{a}_{e,o},\hat{a}_{i,o}, \hat{a}_{e,\Omega}^{\dagger},\hat{a}_{i,\Omega}^{\dagger}]^\text{T}$.
The photon generation rate of the traveling output fields of the electro-optic modulator due to parametric down-conversion $n^\text{out}=\langle a^{\text{out}\dagger}_j(\omega) a^{\text{out}}_j(\omega)\rangle$ can be calculated
\begin{equation}
n^\text{out}=\frac{4C\eta_{j}}{(1-C-\frac{4\omega^2}{\kappa_o\kappa_\Omega})^2+\frac{4\omega^2}{\kappa^2_o\kappa^2_\Omega}(\kappa_o+\kappa_\Omega)^2}.
\label{conversiontraveling} 
\end{equation}
using Eq. (\ref{FDC}), and the bandwidth of the emitted radiation is
\begin{eqnarray}
\text{BW}&= \sqrt{-C-\frac{\kappa^2_o+\kappa_\Omega^2}{2\kappa_o\kappa_\Omega}+\sqrt{(1- C)^2+\left(C+\frac{\kappa^2_o+\kappa_\Omega^2}{2\kappa_o\kappa_\Omega}\right)^2}}&\nonumber\\
&\times\sqrt{\kappa_o\kappa_\Omega},    &       \label{conversionwidth}
\end{eqnarray}
which decreases as a function of $C$ and approaches zero for $C=1$. 

In Fig.~\ref{fig:photonumber}(a) we show the output spectra of the microwave and optical cavities with respect to the response frequency $\omega$ for the experimentally accessible parameters shown in Table~\ref{tab1}
at a cooperativity $C=0.3$ corresponding to a pump power of $P_p=19.2$ $\mu$W.
Even for such low pump powers we obtain readily detectable signal output powers on the order of 1 photon per second per Hertz. Due to the asymmetric waveguide-cavity/resonator coupling $\eta_o\neq\eta_\Omega$ the output photon numbers are not balanced but it is worth noting that the bandwidth is identical even though the dissipation rates $\kappa_\Omega$ and $\kappa_o$ are very different. 
the output spectra 
at an elevated temperature of the cavity baths $T_b=800$~mK related to the thermal photon numbers $\bar{n}^i_{\Omega(o)}=(\exp{(\hbar \Omega(o)/k_B T_b)}-1)^{-1}$. Here we assume a cold waveguide $\bar{n}^e_{\Omega(o)}\sim 0$ which can be realized with superconducting cables connecting to the base temperature of the cryostat. As expected, the output of the microwave cavity increases considerably due to an increase of the modulator thermal noise $\bar n_{\Omega}^i$. 
While the photon occupation of the optical mode $\bar n_{o}^i$ is negligible one can see that the thermal microwave noise leads to parametrically amplified optical noise at the resonator output at elevated temperatures. 

Figure~\ref{fig:photonumber}(b) shows the integrated optical and microwave output photon flux versus multi-photon cooperativity $C$. The photon numbers are increasing with $C$ and diverge abruptly as the cooperativity approaches unity $C\rightarrow 1$. In this limit the system reaches its instability and the linearization approximation used in the Hamiltonian Eq.(\ref{hamsqu}) is not valid anymore. Therefore, for the remainder of the paper we study the generation of microwave-optics entanglement, conversion and quantum state transfer in the parameter range $C<1$.
\section{results}
In this section we verify the generation of microwave-optical two mode squeezing and deterministic entanglement of the output fields in the continuous variable domain. 

\subsection{Two-mode squeezing}
First, we verify the generation of the two-mode squeezing at the outputs of the microwave cavity and optical resonator. For this reason it is convenient to define the field quadratures in terms of the annihilation and creation operators
\begin{equation}\label{quadrature}
\hat{q}_k=\hat{X}_k(0) \mbox{ and } \hat{p}_k=\hat{X}_k(\pi/2),\,\,\,\,\,k=o,\Omega
\end{equation}
where
\begin{equation}
\hat{X}_k(\theta)=\frac{1}{\sqrt{2}} (\hat{A}_k^\text{out}e^{-i\theta}+\hat{A}_k^{\text{out}\dagger} e^{i\theta}).\label{quad}
\end{equation}
These quadratures satisfy the bosonic commutator $[\hat q_k,\hat p_k]=i$ and we define the filtered output operators
\begin{equation}
\hat A_k^\text{out}(\sigma)=\int_{-\infty}^{\infty}d\omega \,\mathrm{f}_k(\omega,\sigma_k)\hat{a}_k^\text{out}(\omega),\label{opintegral}
\end{equation}
where we assume the filter function $\mathrm{f}_k(\omega,\sigma)$ with bandwidth $\sigma_k$ ($k= o, \Omega$) is acting on the output of each cavity. Note that the vacuum noise is $1/2$ for the quadratures defined in Eq. (\ref{quadrature}). 

In order to quantify entanglement, we first determine the covariance matrix (CM) of our system, which can be expressed as
\begin{figure}[t]
	\centering
		\includegraphics[width=0.45\textwidth]{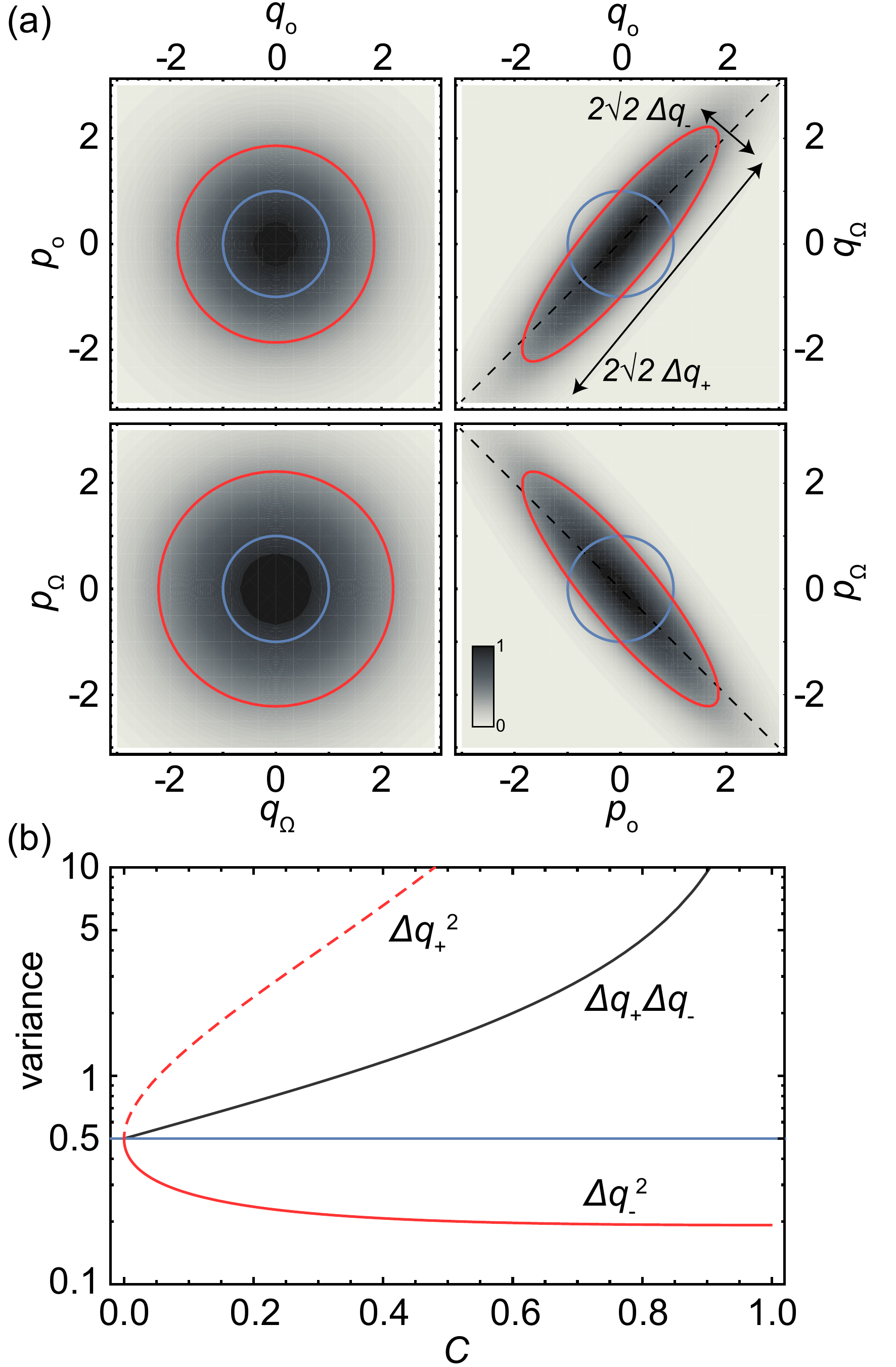}
	\caption{Two-mode squeezing of the electro-optic output fields. 
	(a) Normalized projections of the Wigner function of four output quadrature pairs
	for the same parameters as in Fig.~\ref{fig:photonumber}(a). The solid red line (blue line) indicates a drop by $e^{-1}$ of its (the vacuum state's) maximum. The black dashed-line marks the squeezing angle of $45^\circ$ for an ideal squeezer. The squeezing angles for the asymmetric system in this representation are given by $\pm(90^\circ-\Theta)$. 
	(b) The squeezing $\Delta q_{-}^2$ (solid red line), anti-squeezing parameters $\Delta q_{+}^2$ (dashed red line), their product $\Delta q_{-}\Delta q_{+}$ (black line) and the variance of the vacuum (blue line) as a function of the cooperativity $C$ for the same parameters.}
	\label{QS}
\end{figure}
\begin{equation}
V_{jk}=\frac{1}{2}\langle \Delta \hat{\text{x}}_j \Delta \hat{\text{x}}_k+\Delta \hat{\text{x}}_k\Delta \hat{\text{x}}_j\rangle,
\label{defcov}
\end{equation}
where $\Delta \hat{\text{x}}_k=\hat{\text{x}}_k-\langle \hat{\text{x}}_k\rangle$ and $\hat{\textbf{x}}=[\hat{q}_o,\hat{p}_o,\hat{q}_\Omega,\hat{p}_\Omega]^\text{T}$. Using the scattering matrix defined in Eq.(\ref{FDC}) to calculate the second order moments of the output quadratures Eq. (\ref{quad}) at zero bandwidth $\sigma=0$, we can compute the CM matrix of the system in the steady state
\begin{equation}\label{CM}
\textbf{V} = \begin{bmatrix}
\left(0.5+\frac{4C(1+\bar{n}_\Omega)\eta_o}{(1-C)^2}\right)\textbf{I}&\left(\sqrt{4\eta_o\eta_\Omega C}\frac{(1+C+2\bar{n}_\Omega)}{(1-C)^2}\right)\textbf{Z} \\
\left(\sqrt{4\eta_o\eta_\Omega C}\frac{(1+C+2\bar{n}_\Omega)}{(1-C)^2}\right)\textbf{Z} &\left(0.5+\frac{4(C+\bar{n}_\Omega)\eta_\Omega }{(1-C)^2}\right)\textbf{I} 
\end{bmatrix},
\end{equation}
where $\textbf{I}_{2\times2}$ is the identity matrix, $\textbf{Z}=\mathrm{diag}(1,-1)$,  $\bar{n}_\Omega=\kappa_{i,\Omega} \bar n_{\Omega}^i(T_b)/\kappa_\Omega$ is the microwave thermal mode occupancy. Here we assume a cold waveguide $\bar n_{\Omega(o)}^e=0$ as well as $\bar n_{o}^i=0$.
For $C=0$ the CM Eq.(\ref{CM}) takes on the values of the vacuum noise $\textbf{V}=\textbf{I}_{4\times4}/2$ and the CM diverges at $C=1$. 

The existence of microwave-optical entanglement can be demonstrated using the quasi-probability Wigner function, which can be written in terms of the CM Eq.~(\ref{CM}) and the optical and microwave quadratures $\hat q_k$ and $\hat p_k$
\begin{equation}
W(\textbf{x})=\frac{\exp(-\frac{1}{2}[\textbf{x}\cdot \textbf{V}^{-1}\cdot \textbf{x})]}{\pi^2\sqrt{\text{det}[\textbf{V}]}}\label{wigner}.
\end{equation}
Figure~\ref{QS}(a) shows the Wigner function projected into the 4 different quadratures subspaces $\{p_o,q_o\}$, $\{p_\Omega,q_\Omega\}$, $\{q_\Omega,q_o\}$, and $\{p_\Omega,p_o\}$ where the complementary variables are integrated. As a reference, we also plot the Wigner function of the vacuum state $\textbf{V}=\textbf{I}_{4\times4}/2$ (red circle) corresponding to zero cooperativity $C=0$.  The single-mode projections $\{p_o,q_o\}$ and $\{p_\Omega,q_\Omega\}$ show an increase of the noise fluctuations, indicating the phase-independent amplification of the vacuum noise at the output of each cavity. The  $\{q_\Omega,q_o\}$ and $\{p_\Omega,p_o\}$ projections on the other hand, demonstrate the microwave-optical cross-correlation, originating from the electro-optic interaction, whose fluctuations in specific direction are squeezed below the quantum limit (blue line) and anti-squeezed in the opposite direction. In this plot the red (blue) line indicates a drop by $e^{-1}$ of its maximum for the parameters $C=0.3\, (C=0)$ at $T_b=0$.
Unlike the ideal symmetric two-mode squeezer ($V_{11}=V_{22}= V_{33}=V_{44}$) whose quadrature squeezing appears along diagonal axes with squeezing angle $\pm45^\circ$ (black dashed lines), in general the electro-optic system is an asymmetric squeezer ($V_{11}=V_{22}\neq V_{33}=V_{44}$) if $\eta_o\ne\eta_\Omega$. The squeezing angle is then given by $\tan(2\Theta)=\pm 2V_{13}/|V_{33}-V_{11}|$ and its value is 39.34$^\circ$ for system's parameters in Figure~\ref{QS}(a).

In Fig.~\ref{QS}(b) we show the squeezed $\Delta q_-^2$ and anti-squeezing $\Delta q_+^2$ quadrature variances as well as their product $\Delta q_-\Delta q_+$, which is related to the purity $\mathcal{P}=1/(2 \Delta q_-\Delta q_+)$ of Gaussian states \cite{Paris2003}, as a function of the cooperativity $C$. The variances are given as
\begin{small}
\begin{equation}
\Delta q_{\mp}=\sqrt{\frac{(8C\eta_o+\varepsilon)(8C\eta_\Omega+\varepsilon)/\varepsilon-\Upsilon^2/\varepsilon}{2[\varepsilon+8C\left(\eta_{o,(\Omega)}\sin^2(\Theta)+\eta_{\Omega,(o)}\cos^2(\Theta)\right)\pm\Upsilon\sin(2\Theta)]}},\label{rsq}
\end{equation}\end{small}
with $\Upsilon=4\sqrt{\eta_o\eta_\Omega C}(1+C)$ and $\varepsilon=(1-C)^2$. 
Larger $C$ gives smaller  $\Delta q_-$ (more squeezing) and larger $\Delta q_+$ (more amplification) at the outputs of the cavities. In the ideal case $\eta_o=\eta_\Omega=1$ and for $0<C<1$ the above equation reduces to 
\begin{subequations}
\begin{align}
\Delta q_{-}^2&=\frac{1}{2}\left(\frac{1-\sqrt{C}}{1+\sqrt{C}}\right)^2<\frac{1}{2},\label{qminus}\\
\Delta q_{+}^2&=\frac{1}{2}\left(\frac{1+\sqrt{C}}{1-\sqrt{C}}\right)^2>\frac{1}{2},
\end{align}
\end{subequations}
which satisfies the minimum quadrature uncertainty $\Delta q_-\Delta q_+=1/2$. Moreover, we can define the electro-optic squeezing parameter as $r_\text{EO}=\ln\left(\frac{1+\sqrt{C}}{1-\sqrt{C}}\right)$ for this configuration. Due to the optical and microwave internal losses $\eta_k<1$ ($k=o,\Omega$) the quadrature variances deviate from the uncertainty principle $\Delta q_-\Delta q_+>1/2$ in the proposed device as shown in Fig.~\ref{QS}(b).

\subsection{Microwave-Optical Entanglement}
\begin{figure}[t]
	\centering
		\includegraphics[width=0.4\textwidth]{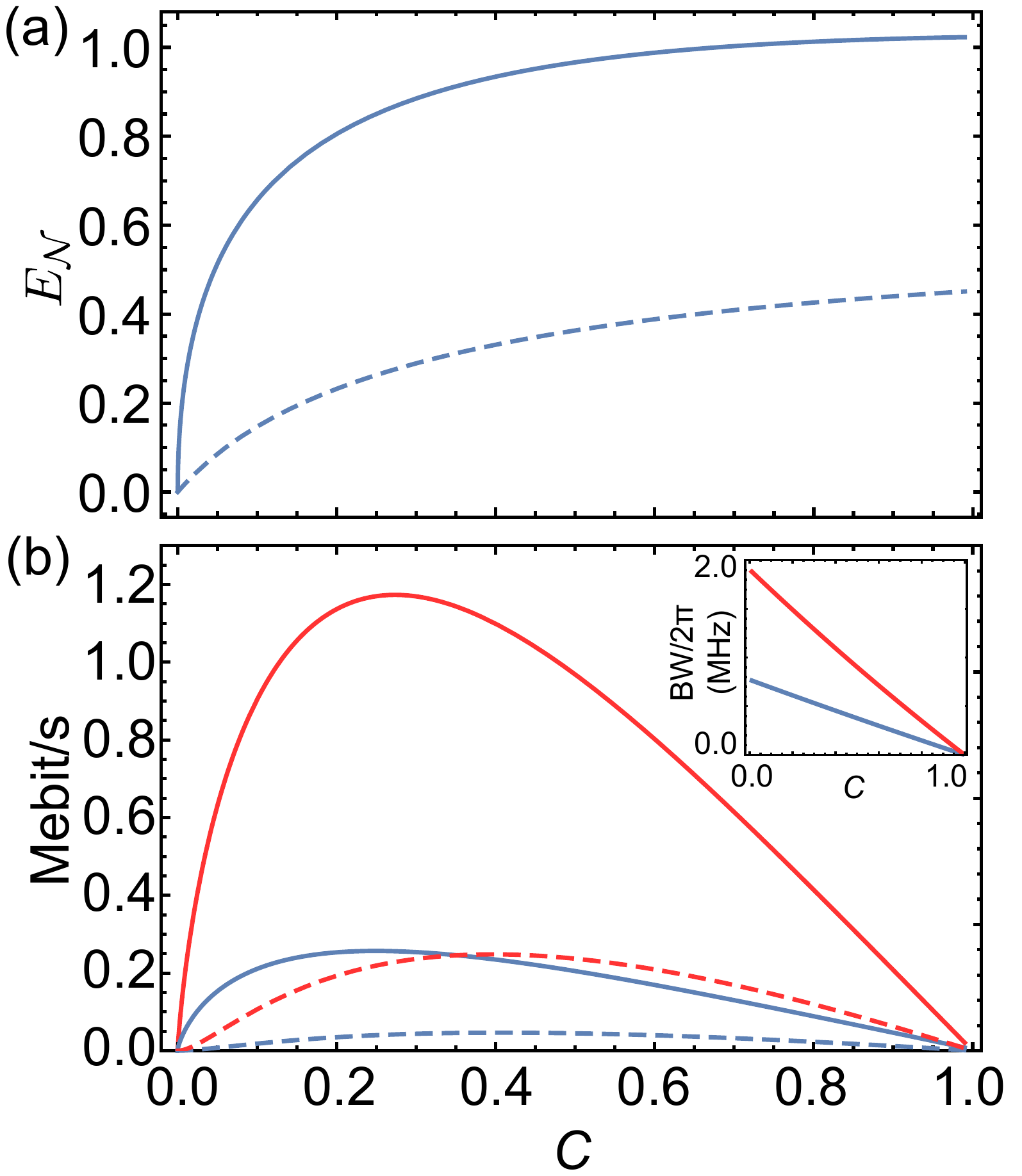}
	\caption{Entanglement and bandwidth of the electro-optic output fields. (a) Microwave-optical entanglement given by the logarithmic negativity $E_\mathcal{N}$ versus cooperativity $C$ at $T_b=10$~mK (solid line) and $T_b=800$~mK (dashed line). (b) The average distribution rate of emitted entangled bits per second at $T_b=10$~mK (solid lines) and $T_b=800$~mK (dashed lines) as a function of cooperativity $C$ for the parameters in Table~\ref{tab1} (blue lines) and stronger optical waveguide coupling, i.e. $\eta_\Omega=\eta_o=0.8$ (red lines). The inset shows the corresponding photon generation bandwidth BW for $\eta_o=0.5$ (blue line) and $\eta_o=0.8$ (red line).
}
	\label{logneg_pic}
\end{figure}

We are interested in the entanglement properties of the radiation leaving the system and we therefore study the bipartite microwave-optical entanglement, which can be quantified using the logarithmic negativity~\cite{Vidal2002, Plenio2005}
\begin{equation}\label{EN}
E_\mathcal{N}=\text{max}[0,-\log_2(2\tilde{d}_-)]. 
\end{equation}
where 
\begin{equation}
 \tilde{d}_-=2^{-1/2}\sqrt{\tilde{\Delta}-\sqrt{\tilde{\Delta}^2-4\text{det}(\textbf{V})}}\label{Logneg},
 \end{equation}
is the smallest symplectic eigenvalue of the partial transpose of the CM Eq.~(\ref{CM}) with $\tilde{\Delta}=V^2_{11}+V^2_{33}+2V^2_{13}$. 
In Fig.~\ref{logneg_pic}(a) we plot $E_\mathcal{N}$ as a function of the cooperativity for two different temperatures 10 mK (solid line) and 800 mK (dashed line). One can see that a significant amount of microwave-optical entanglement is generated $E_\mathcal{N}\sim1$, even for moderate values of $C$, increasing with higher cooperativity and decreasing significantly at elevated bath temperatures $T_b$. In the low temperature limit $\bar n_\Omega\simeq 0$ and for the waveguide coupling matching $\eta:=\eta_o=\eta_\Omega$, the logarithmic negativity (\ref{EN}) reduces to
\begin{equation}
E_\mathcal{N}=-\log_2\left(1-\frac{4\eta\sqrt{C}}{(1+\sqrt{C})^2} \right). \label{lognegeq}
\end{equation}

We also calculate the distribution rate of the entangled fields emitted from the electro-optic system, which is given by
\begin{equation}
\#(\text{ebit/s})=\bar{E}_{F} \cdot \text{BW}/2\pi.
\end{equation}
where we introduce the entanglement of formation 
\begin{equation}
E_F(\rho)=(x_m+0.5)\log_2(x_m+0.5)-(x_m-0.5)\log_2(x_m-0.5) \label{Ef}
\end{equation}
with $x_m=(\tilde{d}_-^2+1/4)/(2\tilde{d}_-)$. 
From the obtained output operators in Eq.~(\ref{opintegral}) with the rectangular filter $\mathrm{f}_k(\omega,\sigma)=\Theta(\text{BW}/2-|\omega|)$ we compute the average CM over the emission bandwidth, which is then used inside Eq.~(\ref{Ef}) returning the average entanglement of formation $\bar{E}_F$.

Figure~\ref{logneg_pic} (b) shows the total emission of entangled radiation as well as the bandwidth of the photon emission as a function of cooperativity $C$. For the considered system parameters given in Table~\ref{tab1} (blue solid line) a maximum value of 0.26 Mebit/s is reached at $C=0.22$ with a photon emission bandwidth of 0.6~MHz at $P_p=14\,\mu$W. The most effective method to increase the BW and entanglement rate is to increase the optical waveguide coupling $\kappa_{e,o}$. The red lines in Fig.~\ref{logneg_pic}(b) show the situation for $\eta_o=0.8$, yielding rates of $>$1~Mebit/s at $\sim2$~MHz bandwidth at $C=0.26$, which would now require a pump power of $P_p=65\,\mu$W, a value that is still feasible at the mixing chamber temperature stage of a dilution refrigerator.  
At significantly elevated bath temperatures of $T_b=800$~mK (dashed lines) the maximum entanglement rates drop by about a factor 5 in both coupling situations. It should be noted that further increasing the coupling to a cold waveguide on the microwave side $\eta_\Omega\simeq1$ or alternatively by finding a way to lower the internal losses of the microwave mode, would result in a significantly smaller effective system temperature. 
Larger waveguide coupling strengths and higher available pump powers at the still stage of a dilution refrigerator together with higher thermal conductivities could result in significantly higher entanglement rates than discussed in this paper which focusses on currently accessible device parameters.
In all cases the entanglement rate approaches zero at $C=1$, following the decrease in photon emission bandwidth, see also Eq. (\ref{conversionwidth}). 

\section{Quantum state transfer}
An important feature of a hybrid quantum network is the ability to transfer quantum states between different nodes. The quality of the state transfer is characterized by the fidelity \cite{braunsteinkimble}  
\begin{equation}
F= \pi\int W_\text{in}(\beta)W_\text{out}(\beta)\text{d}^2\beta,
\end{equation}
where $W_\text{in}$ and $W_\text{out}$ are the initial and final Wigner functions of an unknown arbitrary quantum state before and after the transduction, respectively.  For Gaussian states the fidelity simplifies to \cite{Isar2008}
\begin{equation}
F= \frac{\exp[-(\textbf{x}^\text{out}-\textbf{x}^\text{in})^\text{T}\cdot\textbf{V}_\text{F}^{-1}\cdot(\textbf{x}^\text{out}-\textbf{x}^\text{in})]}{\sqrt{\text{det}(\textbf{V}_\text{F}/2)}} 
\end{equation}
with $\textbf{x}^\text{in}=(q^\text{in}_{o(\Omega)},p^\text{in}_{o(\Omega)})^T$, $\textbf{x}^\text{out}=(q^\text{out}_{o(\Omega)},p^\text{out}_{o(\Omega)})^T$ and $\textbf{V}_\text{F}=2\textbf{V}_\text{in}+2\textbf{V}_\text{out}$, where $\textbf{V}_{\text{in,(out)}}$ are the input and output covariance matrices following the definition given in Eq.~(\ref{defcov}). 

\begin{figure*}[t]
	\centering
		\includegraphics[width=0.85\textwidth]{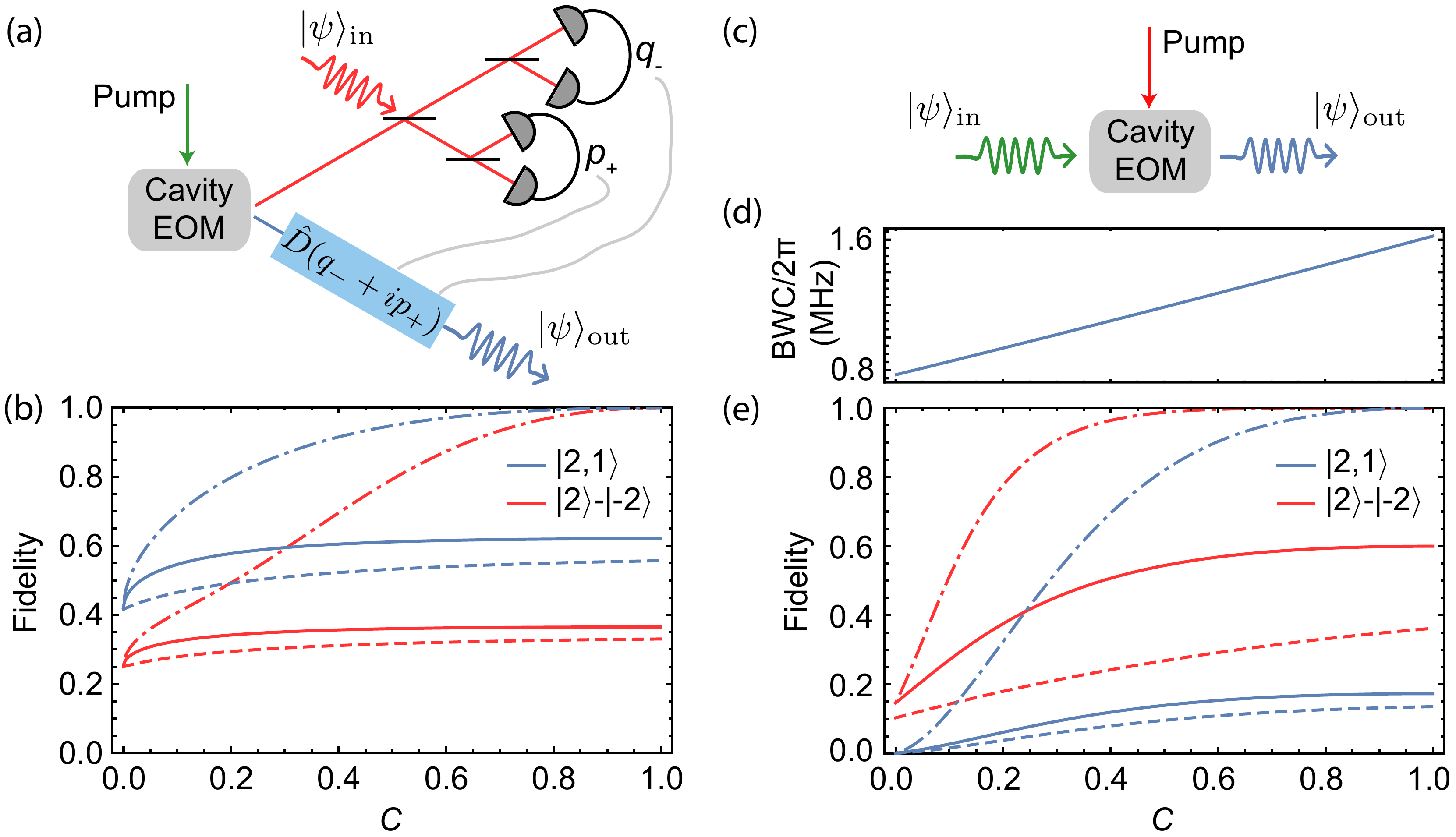}
	\caption{Quantum state transfer. 
	(a) EO teleportation scheme. The sender mixes the unknown optical input state $|\psi\rangle_\text{in}$ with one arm of the EO entanglement source using a 50:50 beam splitter and performs the corresponding Bell-measurements of $q_-$ and $p_+$. This information is sent classically to the microwave receiver, where an appropriate phase space displacement in the second arm of the EO entanglement source is performed to complete the state transfer.
	(b) Calculated fidelity of the teleportation protocol for the coherent squeezed input state $|\psi\rangle_\text{in}=|\alpha=2,r=1\rangle$ (blue lines) and for the cat state $|\psi\rangle_\text{in}=|2\rangle-|-2\rangle$ (red lines) for the experimental parameters outlined in Table~\ref{tab1} (solid lines), at an elevated temperature $T_b=800$~mK (dashed lines) and for the case of a lossless system $\eta_\Omega=\eta_o=1$ (dash-dotted lines).
	(c) Scheme for EO transduction.
	The  EO modulator is coherently pumped on resonance with the lower frequency optical mode \cite{Rueda:16}, allowing for coherent bidirectional conversion between the optical and the microwave modes. 
	(d) Conversion bandwidth as a function of the multi-photon cooperativity $C$ for the experimental parameters outlined in Table~\ref{tab1} .	
	(e) Calculated fidelity of the direct transducer protocol for the coherent squeezed input state $|\psi\rangle_\text{in}=|\alpha=2,r=1\rangle$ (blue lines) and for the cat state $|\psi\rangle_\text{in}=|2\rangle-|-2\rangle$ (red lines) for the experimental parameters outlined in Table~\ref{tab1} (solid lines), at an elevated temperature $T_b=800$~mK (dashed lines) and for the case of a lossless system $\eta_\Omega=\eta_o=1$ (dash-dotted lines). 
	}
	\label{fideltities_1}
\end{figure*}

\subsection{Teleportation}
We propose the bidirectional microwave-to-optical quantum state transfer using the presented EO-device as an EPR source in an unconditional CV teleportation scheme. Assuming the standard Braunstein-Kimble set-up \cite{braunsteinkimble} with ideal Bell measurements and classical information transfer 
as depicted in Fig.~\ref{fideltities_1}(a), the state transfer fidelity for an unknown coherent squeezed state  $|\psi_\text{in}\rangle=|\alpha,r\rangle$ is given by

\begin{equation}
F^\text{G}_\text{TE}(\alpha,r,C,\eta_o,\eta_\Omega)=\left(4\Delta q_-^4+2\Delta q_-^2\cosh(2r)+ 1  \right)^{-1/2}\label{fidelitytele}
\end{equation}
with $\Delta q_-$ explicitly given in Eq.~(\ref{rsq}). In the limit of $\eta_o=\eta_\Omega$ 
the fidelity for pure Gaussian states is reduced to \cite{fiurasek2002}:
\begin{equation}\label{fidsq}
F^\text{G}_\text{TE}(\alpha,r,C,\eta)=\text{Det}[2\textbf{V}_\text{in}+\textbf{ZAZ}+\textbf{B}-\textbf{ZC}-\textbf{C}^T\textbf{Z}^T]^{-1/2},
\end{equation}
where $\textbf{A}=V_{11}\textbf{I}$, $\textbf{B}=V_{33}\textbf{I}$ and $\textbf{C}=V_{13}\textbf{Z}$. The fidelity Eq.~(\ref{fidsq}) can be written in terms of the logarithmic negativity $E_\mathcal{N} $ generated using the EO device
\begin{equation}
F^\text{G}_\text{TE}(\alpha,r,C)=\left(1+2^{1-E_\mathcal{N}(C)}\cosh(2r)+ 2^{-2E_\mathcal{N}(C)}  \right)^{-1/2}\label{FGa}.
\end{equation}

The fidelity in Eq.~(\ref{FGa}) is independent of the coherent state amplitude $\alpha$ due to the assumed ideal measurement of the quadratures $q_-$ and $p_+$, and a lossless classical information transfer in this protocol. The bandwidth of the teleportation is given by the photon emission bandwidth shown in the inset of Fig.~\ref{logneg_pic}(b). 

In Fig.~\ref{fideltities_1}(b) we show the fidelity for the coherent squeezed input state $|\psi_\text{in}\rangle=|\alpha,r \rangle=|2,1\rangle$ as a function of the multi-photon cooperativity $C$ for the system parameters in Table~\ref{tab1} at zero temperature (blue solid line), at 800 mK (blue dashed line) as well as for a lossless system $\eta_o=\eta_{\Omega}=1$ (blue dash-dotted line). The lower bound of the fidelity is set by the classical limit  $F_{\text{TE}}^\text{cl}=e^{-r}/(1+e^{-2r})$ \cite{Owari_2008} valid for non-entangled microwave and optical radiation. The fidelity increases monotonically achieving its maximum value set by the minimum quadrature squeezing of the entanglement source $\Delta q_-^2=0.5-\frac{\eta_o\eta_\Omega}{\eta_o+\eta_\Omega}$ as shown in Fig.~\ref{fideltities_1}(b). An increased temperature leads to a significant reduction of the achievable state transfer fidelity. A fidelity of $\sim 1$ is achieved for a cooperativity close to 1 in the near lossless and perfectly over-coupled case. In this case the system thermalizes with the cold waveguide independent of its internal bath temperature. 

Quantum state teleportation based on EO-entanglement can be used also with non-Gaussian states such as cat states that are readily available in superconducting circuits. Cat states are represented as a quantum superposition of two coherent states in the form $N(|\alpha\rangle+e^{i\phi}|-\alpha\rangle)$ with  $N=\sqrt{2+2\exp(-2\alpha)\cos(\phi)}$. 
The state transfer fidelity using the proposed EO entanglement source is given by \cite{braunsteinkimble}
\begin{equation}
F^\text{cat}_\text{TE}=\frac{1}{1+2\Delta q^2_-}-\frac{1+e^{-4|\alpha|^2}-e^{-4\frac{|\alpha|^2}{1+2\Delta q^2_-}}  -e^{-8\frac{\Delta q^2_-|\alpha|^2}{1+2\Delta q^2_-}}       }{(2+4\Delta q^2_-)(1+e^{-2|\alpha|^2}\cos(\phi))^2}.
\end{equation}
In Fig.~\ref{fideltities_1}(b)  we show the teleportation fidelity of the cat state $|\psi_\text{in}\rangle=|\alpha\rangle-|-\alpha\rangle=|2\rangle-|-2\rangle$ as a function of $C$ 
for the system parameters in Table~\ref{tab1} at zero temperature (red solid line), at 800 mK (red dashed line) as well as for a lossless system $\eta_o=\eta_{\Omega}=1$ (red dash-dotted line) where we consider $\phi=\pi$. We find that the cat state transfer fidelities are lower compared to the coherent squeezed input state over the full range of parameters.
 
\subsection{Conversion}
The EO system can also be used to directly convert the information between microwave and optical photons, schematically shown in Fig.~\ref{fideltities_1}(c). This is achieved by driving the lower frequency optical mode in the same scheme as given in~\cite{Rueda:16,FanZouChengEtAl2018}, changing the nonlinear interaction Hamiltonian into the so called beam splitter Hamiltonian, allowing coherent frequency conversion between the microwave and optical modes following the equations of motion:
\begin{subequations}\label{threemodesUC}
\begin{eqnarray}
\dot{\hat{a}}_{\Omega}&=&-iG \hat{a}_{o}-\frac{\kappa_{\Omega}}{2} \hat{a}_{\Omega}+\sqrt{\kappa_{e,\Omega}}\hat{a}_{e,\Omega}+\sqrt{\kappa_{i,\Omega}}\hat{a}_{i,\Omega},\\
\dot{\hat{a}}_{o}&=&-iG \hat{a}_{\Omega}-\frac{\kappa_{o}}{2} \hat{a}_{o}+\sqrt{\kappa_{e,o}}\hat{a}_{e,o}+\sqrt{\kappa_{i,o}}\hat{a}_{i,o},
\end{eqnarray}
\end{subequations}
Using the input-output theory to calculate the outputs of the optical and microwave modes, we can infer the photon conversion efficiency between the traveling microwave and optical fields 
\cite{tsang_cavity_2011}
\begin{equation}
\frac{\langle a^{\text{out}\dagger}_{o(\Omega)}(\omega) a^{\text{out}}_{o(\Omega)}(\omega)\rangle}{\langle a^{\dagger}_{e,\Omega(o)}(\omega) a_{e,\Omega(o)}(\omega)\rangle}=\frac{4C\eta_\Omega\eta_o}{(1+C-\frac{4\omega^2}{\kappa_o\kappa_\Omega})^2+\frac{4\omega^2}{\kappa^2_o\kappa^2_\Omega}(\kappa_o+\kappa_\Omega)^2},
\label{conversiontravelingUP} 
\end{equation}
over the bandwidth~\cite{RuedaSanchez2018}:
\begin{eqnarray}
\text{BWC}&= \sqrt{C-\frac{\kappa^2_o+\kappa_\Omega^2}{2\kappa_o\kappa_\Omega}+\sqrt{(1+C)^2+\left(C-\frac{\kappa^2_o+\kappa_\Omega^2}{2\kappa_o\kappa_\Omega}\right)^2}}&\nonumber\\
&\times\sqrt{\kappa_o\kappa_\Omega}.    &       \label{conversionwidthup}
\end{eqnarray}

The conversion bandwidth BWC increases with the cooperativity as shown in Fig.~\ref{fideltities_1}(d)  and for the case of rate matching $\kappa_o=\kappa_\Omega=\kappa$  achieves the maximum value of $\sqrt{2}\kappa$ for $C=1$.  For the coherent squeezed state $|\alpha,r\rangle$ the fidelity of the direct state transduction is given by
\begin{equation}
F^\text{G}_\text{tr}(\alpha,r,C)=\frac{\exp\left(-2|\alpha|^2(\epsilon_3-1)^2\left(\frac{\cos(\phi_\alpha) }{V_- }+\frac{\sin(\phi_\alpha)  } {V_+} \right)\right)}{\sqrt{ \frac{\epsilon_2}{2} (1-\epsilon^4_3)+ \epsilon^4_3\left(1  +\frac{\bar{n}_\Omega\left(\epsilon_2+\epsilon_2\epsilon^{-2}_3-2+\frac{\bar{n}_\Omega}{C\eta_o}\right)}{C\eta_o} \right) }}, \label{fidelitytrans}
\end{equation}
where
\begin{equation}
 V_\pm=(1+\epsilon^2_3(e^{\pm2r}-1+2\bar{n}_\Omega/(\eta_o C)),
\end{equation}
$\epsilon_2=1+\cosh(2r)$ and $\epsilon_3=\frac{\sqrt{4\eta_o\eta_\Omega C}}{(1+C)}$. 

Figure \ref{fideltities_1}(e) shows the fidelity of state transfer for the squeezed coherent input state $|\psi_\text{in}\rangle=|2,1\rangle$ as a function of $C$ for the system parameters in Table~\ref{tab1} at zero temperature (blue solid line), at 800 mK (blue dashed line) as well as for a lossless system $\eta_o=\eta_{\Omega}=1$ (blue dash-dotted line). The lower bound of the fidelity ($C=0$) is given by the overlap of the initial state and the vacuum state set by $\frac{2e^{-r-2|\alpha|^2}}{1+e^{-2r}}$. In comparison to the teleportation scheme shown in Fig.~\ref{fideltities_1} (b), the fidelity in direct transduction shown in Figs.~\ref{fideltities_1} (e) is significantly lower for this state. In general for direct conversion the fidelity is strongly dependent on the field amplitude $|\alpha|$, which can be seen from the numerator of Eq.~(\ref{fidelitytrans})
in the case $\eta_{o (\Omega)}<1$. However, it is important to note that many quantum communication protocols work with $|\alpha|\leq1$ \cite{Furusawa706,Wittmann08,cook07}, a regime where both schemes offer more comparable fidelities. 

The direct EO transducer can also be used to convert non-Gaussian cat states between microwave and optical fields. For a real $\alpha$ the fidelity of the conversion is 
\begin{eqnarray}
F^\text{cat}_\text{tr}(\alpha,C)&=&\frac{1}{\epsilon_4(1+\epsilon_5)}[  e^{\frac{-2\alpha^2(1+\epsilon^2_3)^2}{1+\epsilon_5} }  (e^{\frac{8\alpha^2\epsilon_3}{1+\epsilon_5}} +1)       \nonumber\\
&+&2\cos(\phi)( e^{-\frac{2\alpha^2(\epsilon_3^2+\epsilon_5)}{1+\epsilon_5}}  +e^{-\frac{2\alpha^2(1+\epsilon_3^2\epsilon_5)}{1+\epsilon_5}}       ) \nonumber\\
&  +&\cos(2\phi) e^{\frac{-2\alpha^2(\epsilon_5+\epsilon_3)^2}{\epsilon_5(1+\epsilon_5)}}+e^{\frac{-2\alpha^2(\epsilon_3-\epsilon_5)^2}{\epsilon_5(1+\epsilon_5)}} ]   \label{fidelitycattrans}
\end{eqnarray}
with $\epsilon_4=(1+\cos(\phi)e^{-2\alpha^2})(1+\cos(\phi)e^{-2\alpha^2\epsilon^2_3})$ and $\epsilon_5=1+\frac{8\eta_\Omega\bar{n}_\Omega}{(1+C)^2}$, and the lower bound of this fidelity given by $(1+\cos(\phi))/(e^{\alpha^2}+e^{-\alpha^2}\cos(\phi))$. 
In Fig.~\ref{fideltities_1}(e) we plot the conversion fidelity for the cat state $|\psi_\text{in}\rangle=|2\rangle-|-2\rangle$ as a function of $C$ for the system parameters in Table~\ref{tab1} at zero temperature (red solid line), at 800 mK (red dashed line) as well as for a lossless system $\eta_o=\eta_{\Omega}=1$ (red dash-dotted line). We can compare the performance of the two working transduction schemes for the quantum state transfer in electro-optic devices in Figs.~\ref{fideltities_1} (b) and (e). While teleportation performs better for the coherent squeezed state both with and without waveguide coupling losses, for the cat state direct transduction performs better in a broad range $C>0.2$ except for elevated temperatures. It should also be pointed out that the bandwidth of the state transfer is generally higher for direct conversion schemes $\text{BWC}>\text{BW}$ as seen in Figs.~\ref{logneg_pic}(b) and \ref{fideltities_1}(d).

The most efficient electro-optic system yet reported achieved $C=0.075$
with waveguide coupling rates $\eta_o=0.31$ and $\eta_\Omega=0.26$ at an effective temperature of 2.1K~\cite{FanZouChengEtAl2018}. Assuming that the waveguides can be thermalized to low mK temperatures, the fidelities for a state $|\alpha=2,r=1\rangle$ are $10^{-3}$  and 0.41 for direct transduction and teleportation, respectively. On the other hand, the fidelity for an odd cat state with $\alpha=2$ are 0.09 and 0.25 for direct transduction and teleportation schemes, respectively. Our analysis showed that the proposed device should be able to outperform the state of the art with pump powers that are about $10^3$ times lower - a crucial aspect to be able to thermalize the system noise temperature to the cold environmental bath. 

\section{Conclusion}
We have presented an efficient and bright microwave-optical entanglement source based on a triply resonant electro-optic interaction. We proposed a specific device geometry and material system, tested and simulated the most important system parameters and derived the theory describing the physics, entanglement generation and device performance for both teleportation and conversion type quantum state transfer.

The figures of merit for a quantum interface are efficiency and added noise, which both affect the achievable state transfer fidelity. But for any realistic application with finite lifetime qubits, the transducer bandwidth determines if it is of practical use for quantum interconnects. On-chip integrated devices with small mode volume offer higher nonlinear coupling constants $g$ 
\cite{FanZouChengEtAl2018} compared to mm sized systems, but chip-level integration so far comes at the expense of a lower internal optical $Q_{i,o}$ 
\cite{Zhang2017}, because surface qualities routinely achieved with mechanical polishing are difficult to realize in micro-fabrication processes. We have presented a new device geometry that offers the lowest losses without sacrificing coupling and as a result yields high predicted state transfer fidelities at practical bandwidth and realistic optical pump powers.  

Our analysis shows that ultra-low losses, a prerequisite to achieve very strong waveguide over-coupling, turns out to be the most important aspect for any resonant quantum interface to approach the high efficiency and fidelity needed in realistic applications. 
In comparison, increasing waveguide coupling rates requires higher pump powers to achieve the same cooperativity and dissipates more optical energy in the over-coupled regime, which leads to higher thermal bath occupations. 
Our analysis also pointed out the importance of low system temperatures, and mm sized devices not only offer lower optical absorption and scattering losses, which can easily break Cooper pairs in the superconducting microwave cavity, they also offer much larger volume, mass, heat capacity and surface area for effective thermalization to the cold bath in continuous and pulsed driving schemes.

The presented triply resonant modulator offers a very promising way forward in the field of hybrid quantum systems, both when used for entanglement swapping or for direct conversion of quantum states. Experimental tests will show if the proposed scheme can be implemented as expected and tell us more about the important LiNbO$_3$ material parameters and heating rates at millikelvin temperatures. In the context of classical and quantum communication applications, with the above given parameters, our system could also work as an ultra efficient electro-optic modulator with a $V_\pi$ as low as 12.4 mV that can be used for frequency comb generation \cite{Ruedacombs}. Beyond that we also expect applications of microwave-optical entangled fields in the area of radio frequency sensing and low noise detection. 

\section{Acknowledgements}
This work was supported by the Institute of Science and Technology Austria (IST Austria) and the European Research Council under grant agreement number 758053 (ERC StG QUNNECT). S.B. acknowledges support from the Marie Sk\l{}odowska Curie fellowship
number 707438 (MSC-IF SUPEREOM) and J.M.F from the Austrian Science Fund (FWF) through BeyondC (F71), a NOMIS foundation research grant, and the EU's Horizon 2020 research and innovation programme under grant agreement number 732894 (FET Proactive HOT). We thank M. Wulf for manuscript comments and H. Schwefel and C. Marquardt for fruitful discussions.
\section{Author Contributions}
A.R, W.H and J.M.F. conceived the project. Analytical analysis was done by A.R and S.B and FEM simulations by W.H. All authors contributed to the manuscript. S.B. and J.M.F. supervised the project.
\section{additional information}
The authors declare no competing interests.
\bibliographystyle{naturemag_noURL}
\bibliography{FinkGroupBib_v7}

\begin{thebibliography}{10}
\expandafter\ifx\csname url\endcsname\relax
  \def\url#1{\texttt{#1}}\fi
\expandafter\ifx\csname urlprefix\endcsname\relax\def\urlprefix{URL }\fi
\providecommand{\bibinfo}[2]{#2}
\providecommand{\eprint}[2][]{\url{#2}}

\bibitem{schoelkopf_wiring_2008}
\bibinfo{author}{Schoelkopf, R.~J.} \& \bibinfo{author}{Girvin, S.~M.}
\newblock \bibinfo{title}{Wiring up quantum systems}.
\newblock \emph{\bibinfo{journal}{Nature}} \textbf{\bibinfo{volume}{451}},
  \bibinfo{pages}{664--669} (\bibinfo{year}{2008}).

\bibitem{Wendin_2017}
\bibinfo{author}{Wendin, G.}
\newblock \bibinfo{title}{Quantum information processing with superconducting
  circuits: a review}.
\newblock \emph{\bibinfo{journal}{Reports on Progress in Physics}}
  \textbf{\bibinfo{volume}{80}}, \bibinfo{pages}{106001}
  (\bibinfo{year}{2017}).

\bibitem{Kupiers2018}
\bibinfo{author}{Kurpiers, P.} \emph{et~al.}
\newblock \bibinfo{title}{Deterministic quantum state transfer and remote
  entanglement using microwave photons}.
\newblock \emph{\bibinfo{journal}{Nature}} \textbf{\bibinfo{volume}{558}},
  \bibinfo{pages}{264--267} (\bibinfo{year}{2018}).

\bibitem{Chou2018}
\bibinfo{author}{Chou, K.~S.} \emph{et~al.}
\newblock \bibinfo{title}{Deterministic teleportation of a quantum gate between
  two logical qubits}.
\newblock \emph{\bibinfo{journal}{Nature}} \textbf{\bibinfo{volume}{561}},
  \bibinfo{pages}{368--373} (\bibinfo{year}{2018}).

\bibitem{Liao2018}
\bibinfo{author}{Liao, S.-K.} \emph{et~al.}
\newblock \bibinfo{title}{Satellite-relayed intercontinental quantum network}.
\newblock \emph{\bibinfo{journal}{Phys. Rev. Lett.}}
  \textbf{\bibinfo{volume}{120}}, \bibinfo{pages}{030501--}
  (\bibinfo{year}{2018}).

\bibitem{Maring17}
\bibinfo{author}{Maring, N.} \emph{et~al.}
\newblock \bibinfo{title}{Photonic quantum state transfer between a cold atomic
  gas and a crystal}.
\newblock \emph{\bibinfo{journal}{Nature}} \textbf{\bibinfo{volume}{551}},
  \bibinfo{pages}{485 EP --} (\bibinfo{year}{2017}).

\bibitem{Hofheinz2009}
\bibinfo{author}{Hofheinz, M.} \emph{et~al.}
\newblock \bibinfo{title}{Synthesizing arbitrary quantum states in a
  superconducting resonator}.
\newblock \emph{\bibinfo{journal}{Nature}} \textbf{\bibinfo{volume}{459}},
  \bibinfo{pages}{546--549} (\bibinfo{year}{2009}).

\bibitem{Eichler2011a}
\bibinfo{author}{Eichler, C.} \emph{et~al.}
\newblock \bibinfo{title}{Observation of two-mode squeezing in the microwave
  frequency domain}.
\newblock \emph{\bibinfo{journal}{Phys. Rev. Lett.}}
  \textbf{\bibinfo{volume}{107}}, \bibinfo{pages}{113601}
  (\bibinfo{year}{2011}).

\bibitem{Vlastakis2013}
\bibinfo{author}{Vlastakis, B.} \emph{et~al.}
\newblock \bibinfo{title}{Deterministically encoding quantum information using
  100-photon schrödinger cat states}.
\newblock \emph{\bibinfo{journal}{Science}} \textbf{\bibinfo{volume}{342}},
  \bibinfo{pages}{607--} (\bibinfo{year}{2013}).

\bibitem{Braunsteinvanloock}
\bibinfo{author}{Braunstein, S.~L.} \& \bibinfo{author}{van Loock, P.}
\newblock \bibinfo{title}{Quantum information with continuous variables}.
\newblock \emph{\bibinfo{journal}{Rev. Mod. Phys.}}
  \textbf{\bibinfo{volume}{77}}, \bibinfo{pages}{513--577}
  (\bibinfo{year}{2005}).

\bibitem{weedbrook2012}
\bibinfo{author}{Weedbrook, C.} \emph{et~al.}
\newblock \bibinfo{title}{Gaussian quantum information}.
\newblock \emph{\bibinfo{journal}{Rev. Mod. Phys.}}
  \textbf{\bibinfo{volume}{84}}, \bibinfo{pages}{621--669}
  (\bibinfo{year}{2012}).

\bibitem{Furusawa706}
\bibinfo{author}{Furusawa, A.} \emph{et~al.}
\newblock \bibinfo{title}{Unconditional quantum teleportation}.
\newblock \emph{\bibinfo{journal}{Science}} \textbf{\bibinfo{volume}{282}},
  \bibinfo{pages}{706--709} (\bibinfo{year}{1998}).

\bibitem{Lee330}
\bibinfo{author}{Lee, N.} \emph{et~al.}
\newblock \bibinfo{title}{Teleportation of nonclassical wave packets of light}.
\newblock \emph{\bibinfo{journal}{Science}} \textbf{\bibinfo{volume}{332}},
  \bibinfo{pages}{330--333} (\bibinfo{year}{2011}).

\bibitem{Laurat03}
\bibinfo{author}{Laurat, J.}, \bibinfo{author}{Coudreau, T.},
  \bibinfo{author}{Treps, N.}, \bibinfo{author}{Ma\^{\i}tre, A.} \&
  \bibinfo{author}{Fabre, C.}
\newblock \bibinfo{title}{Conditional preparation of a quantum state in the
  continuous variable regime: Generation of a sub-poissonian state from twin
  beams}.
\newblock \emph{\bibinfo{journal}{Phys. Rev. Lett.}}
  \textbf{\bibinfo{volume}{91}}, \bibinfo{pages}{213601}
  (\bibinfo{year}{2003}).

\bibitem{pogorzalek2019}
\bibinfo{author}{{Pogorzalek}, S.} \emph{et~al.}
\newblock \bibinfo{title}{{Secure quantum remote state preparation of squeezed
  microwave states}}.
\newblock \emph{\bibinfo{journal}{arXiv:1902.00453}}  (\bibinfo{year}{2019}).

\bibitem{Andrews2014}
\bibinfo{author}{Andrews, R.~W.} \emph{et~al.}
\newblock \bibinfo{title}{Bidirectional and efficient conversion between
  microwave and optical light}.
\newblock \emph{\bibinfo{journal}{Nature Physics}}
  \textbf{\bibinfo{volume}{10}}, \bibinfo{pages}{321--326}
  (\bibinfo{year}{2014}).

\bibitem{higgin}
\bibinfo{author}{Higginbotham, A.~P.} \emph{et~al.}
\newblock \bibinfo{title}{Harnessing electro-optic correlations in an efficient
  mechanical converter}.
\newblock \emph{\bibinfo{journal}{Nature Physics}}
  \textbf{\bibinfo{volume}{14}}, \bibinfo{pages}{1038--1042}
  (\bibinfo{year}{2018}).

\bibitem{Barzanjeh2019}
\bibinfo{author}{Barzanjeh, S.} \emph{et~al.}
\newblock \bibinfo{title}{Stationary entangled radiation from micromechanical
  motion}.
\newblock \emph{\bibinfo{journal}{Nature}} \textbf{\bibinfo{volume}{570}},
  \bibinfo{pages}{480--483} (\bibinfo{year}{2019}).

\bibitem{Genes2008b}
\bibinfo{author}{Genes, C.}, \bibinfo{author}{Mari, A.},
  \bibinfo{author}{Tombesi, P.} \& \bibinfo{author}{Vitali, D.}
\newblock \bibinfo{title}{Robust entanglement of a micromechanical resonator
  with output optical fields}.
\newblock \emph{\bibinfo{journal}{Phys.\ Rev.\ A}}
  \textbf{\bibinfo{volume}{78}}, \bibinfo{pages}{032316}
  (\bibinfo{year}{2008}).

\bibitem{Stannigel2010}
\bibinfo{author}{Stannigel, K.}, \bibinfo{author}{Rabl, P.},
  \bibinfo{author}{S\o{}rensen, A.~S.}, \bibinfo{author}{Zoller, P.} \&
  \bibinfo{author}{Lukin, M.~D.}
\newblock \bibinfo{title}{Optomechanical transducers for long-distance quantum
  communication}.
\newblock \emph{\bibinfo{journal}{Phys. Rev. Lett.}}
  \textbf{\bibinfo{volume}{105}}, \bibinfo{pages}{220501}
  (\bibinfo{year}{2010}).

\bibitem{Barzanjeh2011a}
\bibinfo{author}{Barzanjeh, S.}, \bibinfo{author}{Vitali, D.},
  \bibinfo{author}{Tombesi, P.} \& \bibinfo{author}{Milburn, G.~J.}
\newblock \bibinfo{title}{Entangling optical and microwave cavity modes by
  means of a nanomechanical resonator}.
\newblock \emph{\bibinfo{journal}{Phys. Rev. A}} \textbf{\bibinfo{volume}{84}},
  \bibinfo{pages}{042342} (\bibinfo{year}{2011}).

\bibitem{shab1}
\bibinfo{author}{Barzanjeh, S.}, \bibinfo{author}{Abdi, M.},
  \bibinfo{author}{Milburn, G.~J.}, \bibinfo{author}{Tombesi, P.} \&
  \bibinfo{author}{Vitali, D.}
\newblock \bibinfo{title}{Reversible optical-to-microwave quantum interface}.
\newblock \emph{\bibinfo{journal}{Phys. Rev. Lett.}}
  \textbf{\bibinfo{volume}{109}}, \bibinfo{pages}{130503}
  (\bibinfo{year}{2012}).

\bibitem{Wang2013}
\bibinfo{author}{Wang, Y.-D.} \& \bibinfo{author}{Clerk, A.~A.}
\newblock \bibinfo{title}{Reservoir-engineered entanglement in optomechanical
  systems}.
\newblock \emph{\bibinfo{journal}{Phys. Rev. Lett.}}
  \textbf{\bibinfo{volume}{110}}, \bibinfo{pages}{253601--}
  (\bibinfo{year}{2013}).

\bibitem{Tian2013}
\bibinfo{author}{Tian, L.}
\newblock \bibinfo{title}{Robust photon entanglement via quantum interference
  in optomechanical interfaces}.
\newblock \emph{\bibinfo{journal}{Phys. Rev. Lett.}}
  \textbf{\bibinfo{volume}{110}}, \bibinfo{pages}{233602}
  (\bibinfo{year}{2013}).

\bibitem{Zhong2019}
\bibinfo{author}{{Zhong}, C.} \emph{et~al.}
\newblock \bibinfo{title}{{Heralded Generation and Detection of Entangled
  Microwave--Optical Photon Pairs}}.
\newblock \emph{\bibinfo{journal}{arXiv e-prints}}
  \bibinfo{pages}{arXiv:1901.08228} (\bibinfo{year}{2019}).
\newblock arXiv:\eprint{1901.08228}.

\bibitem{Bochmann2013}
\bibinfo{author}{Bochmann, J.}, \bibinfo{author}{Vainsencher, A.},
  \bibinfo{author}{Awschalom, D.~D.} \& \bibinfo{author}{Cleland, A.~N.}
\newblock \bibinfo{title}{Nanomechanical coupling between microwave and optical
  photons}.
\newblock \emph{\bibinfo{journal}{Nature Physics}}
  \textbf{\bibinfo{volume}{9}}, \bibinfo{pages}{712--716}
  (\bibinfo{year}{2013}).

\bibitem{Forsch2018}
\bibinfo{author}{Forsch, M.} \emph{et~al.}
\newblock \bibinfo{title}{Microwave-to-optics conversion using a mechanical
  oscillator in its quantum groundstate}.
\newblock \emph{\bibinfo{journal}{arXiv:1812.07588v1}}  (\bibinfo{year}{2018}).

\bibitem{Shao2019}
\bibinfo{author}{Shao, L.} \emph{et~al.}
\newblock \bibinfo{title}{Microwave-to-optical conversion using lithium niobate
  thin-film acoustic resonators}.
\newblock \emph{\bibinfo{journal}{arXiv:1907.08593}}  (\bibinfo{year}{2019}).

\bibitem{hisatomi}
\bibinfo{author}{Hisatomi, R.} \emph{et~al.}
\newblock \bibinfo{title}{Bidirectional conversion between microwave and light
  via ferromagnetic magnons}.
\newblock \emph{\bibinfo{journal}{Phys. Rev. B}} \textbf{\bibinfo{volume}{93}},
  \bibinfo{pages}{174427} (\bibinfo{year}{2016}).

\bibitem{Matsko2007}
\bibinfo{author}{Matsko, A.~B.}, \bibinfo{author}{Savchenkov, A.~A.},
  \bibinfo{author}{Ilchenko, V.~S.}, \bibinfo{author}{Seidel, D.} \&
  \bibinfo{author}{Maleki, L.}
\newblock \bibinfo{title}{On fundamental quantum noises of whispering gallery
  mode electro-optic modulators}.
\newblock \emph{\bibinfo{journal}{Opt. Express}} \textbf{\bibinfo{volume}{15}},
  \bibinfo{pages}{17401--17409} (\bibinfo{year}{2007}).

\bibitem{tsang_cavity_2010}
\bibinfo{author}{Tsang, M.}
\newblock \bibinfo{title}{Cavity quantum electro-optics}.
\newblock \emph{\bibinfo{journal}{Physical Review A}}
  \textbf{\bibinfo{volume}{81}}, \bibinfo{pages}{063837}
  (\bibinfo{year}{2010}).

\bibitem{tsang_cavity_2011}
\bibinfo{author}{Tsang, M.}
\newblock \bibinfo{title}{Cavity quantum electro-optics. {II}. {Input}-output
  relations between traveling optical and microwave fields}.
\newblock \emph{\bibinfo{journal}{Physical Review A}}
  \textbf{\bibinfo{volume}{84}}, \bibinfo{pages}{043845}
  (\bibinfo{year}{2011}).

\bibitem{Javerzac-Galy2016}
\bibinfo{author}{Javerzac-Galy, C.} \emph{et~al.}
\newblock \bibinfo{title}{On-chip microwave-to-optical quantum coherent
  converter based on a superconducting resonator coupled to an electro-optic
  microresonator}.
\newblock \emph{\bibinfo{journal}{Phys. Rev. A}} \textbf{\bibinfo{volume}{94}},
  \bibinfo{pages}{053815} (\bibinfo{year}{2016}).

\bibitem{Soltani2017}
\bibinfo{author}{Soltani, M.} \emph{et~al.}
\newblock \bibinfo{title}{Efficient quantum microwave-to-optical conversion
  using electro-optic nanophotonic coupled resonators}.
\newblock \emph{\bibinfo{journal}{Phys. Rev. A}} \textbf{\bibinfo{volume}{96}},
  \bibinfo{pages}{043808--} (\bibinfo{year}{2017}).

\bibitem{cohen_microphotonic_2001-1}
\bibinfo{author}{Cohen, D.}, \bibinfo{author}{Hossein-Zadeh, M.} \&
  \bibinfo{author}{Levi, A.}
\newblock \bibinfo{title}{Microphotonic modulator for microwave receiver}.
\newblock \emph{\bibinfo{journal}{Electronics Letters}}
  \textbf{\bibinfo{volume}{37}}, \bibinfo{pages}{300--301}
  (\bibinfo{year}{2001}).

\bibitem{ilchenko_whispering-gallery-mode_2003}
\bibinfo{author}{Ilchenko, V.~S.}, \bibinfo{author}{Savchenkov, A.~A.},
  \bibinfo{author}{Matsko, A.~B.} \& \bibinfo{author}{Maleki, L.}
\newblock \bibinfo{title}{Whispering-gallery-mode electro-optic modulator and
  photonic microwave receiver}.
\newblock \emph{\bibinfo{journal}{Journal of the Optical Society of America B}}
  \textbf{\bibinfo{volume}{20}}, \bibinfo{pages}{333--342}
  (\bibinfo{year}{2003}).

\bibitem{strekalov_efficient_2009}
\bibinfo{author}{Strekalov, D.~V.}, \bibinfo{author}{Savchenkov, A.~A.},
  \bibinfo{author}{Matsko, A.~B.} \& \bibinfo{author}{Yu, N.}
\newblock \bibinfo{title}{Efficient upconversion of subterahertz radiation in a
  high-{Q} whispering gallery resonator}.
\newblock \emph{\bibinfo{journal}{Optics Letters}}
  \textbf{\bibinfo{volume}{34}}, \bibinfo{pages}{713--715}
  (\bibinfo{year}{2009}).

\bibitem{strekalov_microwave_2009}
\bibinfo{author}{Strekalov, D.~V.} \emph{et~al.}
\newblock \bibinfo{title}{Microwave whispering-gallery resonator for efficient
  optical up-conversion}.
\newblock \emph{\bibinfo{journal}{Physical Review A.}}
  \textbf{\bibinfo{volume}{80}}, \bibinfo{pages}{033810--5}
  (\bibinfo{year}{2009}).

\bibitem{botello_sensitivity_2018}
\bibinfo{author}{Botello, G. S.-a.} \emph{et~al.}
\newblock \bibinfo{title}{Sensitivity limits of millimeter-wave photonic
  radiometers based on efficient electro-optic upconverters}.
\newblock \emph{\bibinfo{journal}{Optica}} \textbf{\bibinfo{volume}{5}},
  \bibinfo{pages}{1210--1219} (\bibinfo{year}{2018}).

\bibitem{Rueda:16}
\bibinfo{author}{Rueda, A.} \emph{et~al.}
\newblock \bibinfo{title}{Efficient microwave to optical photon conversion: an
  electro-optical realization}.
\newblock \emph{\bibinfo{journal}{Optica}} \textbf{\bibinfo{volume}{3}},
  \bibinfo{pages}{597--604} (\bibinfo{year}{2016}).

\bibitem{FanZouChengEtAl2018}
\bibinfo{author}{Fan, L.} \emph{et~al.}
\newblock \bibinfo{title}{Superconducting cavity electro-optics: A platform for
  coherent photon conversion between superconducting and photonic circuits}.
\newblock \emph{\bibinfo{journal}{Science Advances}}
  \textbf{\bibinfo{volume}{4}} (\bibinfo{year}{2018}).

\bibitem{reviewstrekalov}
\bibinfo{author}{Strekalov, D.~V.}, \bibinfo{author}{Marquardt, C.},
  \bibinfo{author}{Matsko, A.~B.}, \bibinfo{author}{Schwefel, H. G.~L.} \&
  \bibinfo{author}{Leuchs, G.}
\newblock \bibinfo{title}{{Nonlinear and quantum optics with whispering gallery
  resonators}}.
\newblock \emph{\bibinfo{journal}{Journal of Optics}}
  \textbf{\bibinfo{volume}{18}}, \bibinfo{pages}{123002}
  (\bibinfo{year}{2016}).

\bibitem{Muralidharan2016}
\bibinfo{author}{Muralidharan, S.} \emph{et~al.}
\newblock \bibinfo{title}{Optimal architectures for long distance quantum
  communication}.
\newblock \emph{\bibinfo{journal}{Scientific Reports}}
  \textbf{\bibinfo{volume}{6}}, \bibinfo{pages}{20463--}
  (\bibinfo{year}{2016}).

\bibitem{Leidinger:15}
\bibinfo{author}{Leidinger, M.} \emph{et~al.}
\newblock \bibinfo{title}{Comparative study on three highly sensitive
  absorption measurement techniques characterizing lithium niobate over its
  entire transparent spectral range}.
\newblock \emph{\bibinfo{journal}{Opt. Express}} \textbf{\bibinfo{volume}{23}},
  \bibinfo{pages}{21690--21705} (\bibinfo{year}{2015}).

\bibitem{RuedaSanchez2018}
\bibinfo{author}{Sanchez, A. R.~R.}
\newblock \emph{\bibinfo{title}{Resonant Electrooptics}}.
\newblock \bibinfo{type}{doctoralthesis},
  \bibinfo{school}{Friedrich-Alexander-Universit{\"a}t Erlangen-N{\"u}rnberg
  (FAU)} (\bibinfo{year}{2018}).

\bibitem{goryachev_single-photon_2015}
\bibinfo{author}{Goryachev, M.}, \bibinfo{author}{Kostylev, N.} \&
  \bibinfo{author}{Tobar, M.~E.}
\newblock \bibinfo{title}{Single-photon level study of microwave properties of
  lithium niobate at millikelvin temperatures}.
\newblock \emph{\bibinfo{journal}{Physical Review B}}
  \textbf{\bibinfo{volume}{92}}, \bibinfo{pages}{060406}
  (\bibinfo{year}{2015}).

\bibitem{Weis1985}
\bibinfo{author}{Weis, R.~S.} \& \bibinfo{author}{Gaylord, T.~K.}
\newblock \bibinfo{title}{Lithium niobate: Summary of physical properties and
  crystal structure}.
\newblock \emph{\bibinfo{journal}{Applied Physics A}}
  \textbf{\bibinfo{volume}{37}}, \bibinfo{pages}{191--203}
  (\bibinfo{year}{1985}).

\bibitem{wong2002properties}
\bibinfo{author}{Wong, K.}, \bibinfo{author}{of~Electrical~Engineers, I.} \&
  \bibinfo{author}{service), I.~I.}
\newblock \emph{\bibinfo{title}{Properties of Lithium Niobate}}.
\newblock EMIS datareviews series (\bibinfo{publisher}{INSPEC/Institution of
  Electrical Engineers}, \bibinfo{year}{2002}).

\bibitem{Aluimag}
\bibinfo{author}{McPeak, K.~M.} \emph{et~al.}
\newblock \bibinfo{title}{Plasmonic films can easily be better: Rules and
  recipes}.
\newblock \emph{\bibinfo{journal}{ACS Photonics}} \textbf{\bibinfo{volume}{2}},
  \bibinfo{pages}{326--333} (\bibinfo{year}{2015}).
\newblock \bibinfo{note}{PMID: 25950012},
  arXiv:\eprint{https://doi.org/10.1021/ph5004237}.

\bibitem{Rakic:95}
\bibinfo{author}{Raki\'{c}, A.~D.}
\newblock \bibinfo{title}{Algorithm for the determination of intrinsic optical
  constants of metal films: application to aluminum}.
\newblock \emph{\bibinfo{journal}{Appl. Opt.}} \textbf{\bibinfo{volume}{34}},
  \bibinfo{pages}{4755--4767} (\bibinfo{year}{1995}).

\bibitem{Nguyen2013}
\bibinfo{author}{Nguyen, D.~T.} \emph{et~al.}
\newblock \bibinfo{title}{Ultrahigh q-frequency product for optomechanical disk
  resonators with a mechanical shield}.
\newblock \emph{\bibinfo{journal}{Appl. Phys. Lett.}}
  \textbf{\bibinfo{volume}{103}}, \bibinfo{pages}{241112}
  (\bibinfo{year}{2013}).

\bibitem{Brecht16}
\bibinfo{author}{Brecht, T.} \emph{et~al.}
\newblock \bibinfo{title}{Multilayer microwave integrated quantum circuits for
  scalable quantum computing}.
\newblock \emph{\bibinfo{journal}{Npj Quantum Information}}
  \textbf{\bibinfo{volume}{2}}, \bibinfo{pages}{16002 EP --}
  (\bibinfo{year}{2016}).

\bibitem{werner11}
\bibinfo{author}{Wenner, J.} \emph{et~al.}
\newblock \bibinfo{title}{Surface loss simulations of superconducting coplanar
  waveguide resonators}.
\newblock \emph{\bibinfo{journal}{Applied Physics Letters}}
  \textbf{\bibinfo{volume}{99}}, \bibinfo{pages}{113513}
  (\bibinfo{year}{2011}).

\bibitem{Gardiner2004}
\bibinfo{author}{Gardiner, C.~W.} \& \bibinfo{author}{Zoller, P.}
\newblock \emph{\bibinfo{title}{Quantum Noise}} (\bibinfo{publisher}{Springer
  Series in Synergetics}, \bibinfo{year}{2004}).

\bibitem{Paris2003}
\bibinfo{author}{Paris, M. G.~A.}, \bibinfo{author}{Illuminati, F.},
  \bibinfo{author}{Serafini, A.} \& \bibinfo{author}{De~Siena, S.}
\newblock \bibinfo{title}{Purity of gaussian states: Measurement schemes and
  time evolution in noisy channels}.
\newblock \emph{\bibinfo{journal}{Phys. Rev. A}} \textbf{\bibinfo{volume}{68}},
  \bibinfo{pages}{012314--} (\bibinfo{year}{2003}).

\bibitem{Vidal2002}
\bibinfo{author}{Vidal, G.} \& \bibinfo{author}{Werner, R.~F.}
\newblock \bibinfo{title}{Computable measure of entanglement}.
\newblock \emph{\bibinfo{journal}{Phys. Rev. A}} \textbf{\bibinfo{volume}{65}},
  \bibinfo{pages}{032314} (\bibinfo{year}{2002}).

\bibitem{Plenio2005}
\bibinfo{author}{Plenio, M.~B.}
\newblock \bibinfo{title}{Logarithmic negativity: A full entanglement monotone
  that is not convex}.
\newblock \emph{\bibinfo{journal}{Phys. Rev. Lett.}}
  \textbf{\bibinfo{volume}{95}}, \bibinfo{pages}{090503}
  (\bibinfo{year}{2005}).

\bibitem{braunsteinkimble}
\bibinfo{author}{Braunstein, S.~L.} \& \bibinfo{author}{Kimble, H.~J.}
\newblock \bibinfo{title}{Teleportation of continuous quantum variables}.
\newblock \emph{\bibinfo{journal}{Phys. Rev. Lett.}}
  \textbf{\bibinfo{volume}{80}}, \bibinfo{pages}{869--872}
  (\bibinfo{year}{1998}).

\bibitem{Isar2008}
\bibinfo{author}{Isar, A.}
\newblock \bibinfo{title}{Quantum fidelity for {Gauss}ian states describing the
  evolution of open systems}.
\newblock \emph{\bibinfo{journal}{The European Physical Journal Special
  Topics}} \textbf{\bibinfo{volume}{160}}, \bibinfo{pages}{225--234}
  (\bibinfo{year}{2008}).

\bibitem{fiurasek2002}
\bibinfo{author}{Fiur\'a\ifmmode~\check{s}\else \v{s}\fi{}ek, J.}
\newblock \bibinfo{title}{Improving the fidelity of continuous-variable
  teleportation via local operations}.
\newblock \emph{\bibinfo{journal}{Phys. Rev. A}} \textbf{\bibinfo{volume}{66}},
  \bibinfo{pages}{012304} (\bibinfo{year}{2002}).

\bibitem{Owari_2008}
\bibinfo{author}{Owari, M.}, \bibinfo{author}{Plenio, M.~B.},
  \bibinfo{author}{Polzik, E.~S.}, \bibinfo{author}{Serafini, A.} \&
  \bibinfo{author}{Wolf, M.~M.}
\newblock \bibinfo{title}{Squeezing the limit: quantum benchmarks for the
  teleportation and storage of squeezed states}.
\newblock \emph{\bibinfo{journal}{New Journal of Physics}}
  \textbf{\bibinfo{volume}{10}}, \bibinfo{pages}{113014}
  (\bibinfo{year}{2008}).

\bibitem{Wittmann08}
\bibinfo{author}{Wittmann, C.} \emph{et~al.}
\newblock \bibinfo{title}{Demonstration of near-optimal discrimination of
  optical coherent states}.
\newblock \emph{\bibinfo{journal}{Phys. Rev. Lett.}}
  \textbf{\bibinfo{volume}{101}}, \bibinfo{pages}{210501}
  (\bibinfo{year}{2008}).

\bibitem{cook07}
\bibinfo{author}{Cook, R.~L.}, \bibinfo{author}{Martin, P.~J.} \&
  \bibinfo{author}{Geremia, J.~M.}
\newblock \bibinfo{title}{Optical coherent state discrimination using a
  closed-loop quantum measurement}.
\newblock \emph{\bibinfo{journal}{Nature}} \textbf{\bibinfo{volume}{446}},
  \bibinfo{pages}{774 EP --} (\bibinfo{year}{2007}).

\bibitem{Zhang2017}
\bibinfo{author}{Zhang, M.}, \bibinfo{author}{Wang, C.},
  \bibinfo{author}{Cheng, R.}, \bibinfo{author}{Shams-Ansari, A.} \&
  \bibinfo{author}{Lončar, M.}
\newblock \bibinfo{title}{Monolithic ultra-high-q lithium niobate microring
  resonator}.
\newblock \emph{\bibinfo{journal}{Optica}} \textbf{\bibinfo{volume}{4}},
  \bibinfo{pages}{1536--1537} (\bibinfo{year}{2017}).

\bibitem{Ruedacombs}
\bibinfo{author}{Rueda, A.}, \bibinfo{author}{Sedlmeir, F.},
  \bibinfo{author}{Kumari, M.}, \bibinfo{author}{Leuchs, G.} \&
  \bibinfo{author}{Schwefel, H. G.~L.}
\newblock \bibinfo{title}{Resonant electro-optic frequency comb}.
\newblock \emph{\bibinfo{journal}{Nature}} \textbf{\bibinfo{volume}{568}},
  \bibinfo{pages}{378--381} (\bibinfo{year}{2019}).

\end{thebibliography}
\end{document}